\documentclass[10pt,journal]{IEEEtran}

\makeatletter
\def\ps@headings{%
\def\@oddhead{\mbox{}\scriptsize\rightmark \hfil \thepage}%
\def\@evenhead{\scriptsize\thepage \hfil \leftmark\mbox{}}%
\def\@oddfoot{}%
\def\@evenfoot{}}
\makeatother 
\usepackage{pdfpages}
\usepackage{epstopdf}
\usepackage{booktabs}
\usepackage{caption}
\usepackage{setspace}
\usepackage{diagbox}
\usepackage{tipa}
\usepackage{stmaryrd}
\usepackage{pifont}
\usepackage{amssymb}
\usepackage{amsmath,amsthm}
\usepackage{cite}
\usepackage{amsfonts}
\usepackage{txfonts}
\usepackage{amsmath}
\usepackage{latexsym}
\usepackage{cases}
\usepackage{subfigure}
\usepackage{graphicx}
\usepackage{algorithm}
\usepackage{algorithmic}
\usepackage{bbding}
\usepackage{times}
\usepackage{array}
\usepackage{color}
\usepackage{mathrsfs}
\usepackage{multirow}
\renewcommand{\arraystretch}{1.7}

\def\comment#1{}
\begin{document}

\title{VRConvMF: Visual Recurrent Convolutional Matrix Factorization for Movie Recommendation}

\author{Zhu Wang$^{*}$, Honglong Chen$^{*}$, Zhe Li$^{*}$, Kai Lin$^{*}$, Nan Jiang$^{\S}$, and Feng Xia$^{\P}$\\
$^*$College of Information and Control Engineering, China
University of Petroleum, Qingdao, China\\
$^{\S}$College of Information Engineering, East China Jiaotong University, Nanchang, China\\
$^{\P}$School of Engineering, IT and Physical Sciences, Federation University Australia, Ballarat, Australia
\thanks{This work was supported by NSFC
under Grant 61772551, Grant 62111530052 and the Major Scientific and Technological Projects
of CNPC under Grant ZD2019-183-003.}}

\IEEEcompsoctitleabstractindextext{

\begin{abstract}
  Sparsity of user-to-item rating data becomes one of challenging issues in the recommender systems, which severely deteriorates the recommendation performance. Fortunately, context-aware recommender systems can alleviate the sparsity problem by making use of some auxiliary information, such as the information of both the users and items. In particular, the visual information of items, such as the movie poster, can be considered as the supplement for item description documents, which helps to obtain more item features. In this paper, we focus on movie recommender system and propose a probabilistic matrix factorization based recommendation scheme called \textit{visual recurrent convolutional matrix factorization} (VRConvMF), which utilizes the textual and multi-level visual features extracted from the descriptive texts and posters respectively. We implement the proposed VRConvMF and conduct extensive experiments on three commonly used real world datasets to validate its effectiveness. The experimental results illustrate that the proposed VRConvMF outperforms the existing schemes.
\end{abstract}

}
\maketitle
\IEEEdisplaynotcompsoctitleabstractindextext
\IEEEpeerreviewmaketitle
\renewcommand{\thefootnote}{\fnsymbol{footnote}}
\setcounter{footnote}{0}
\footnotetext{Corresponding author: Honglong Chen. Email: chenhl@upc.edu.cn.}

\section{Introduction}\label{sec1}
The rapid development of the Internet of Things (IoTs)~\cite{Aixin:2021,Lin:2021} leads to the explosive growth of the number of items and users. Therefore, the sparseness of the user-to-item rating matrix in e-commerce becomes more and more serious, resulting in deteriorating the rating prediction precision of conventional collaborative filtering algorithms~\cite{G.Adomavicius}. For instance, in the commonly used MovieLens-10m dataset, the average number of movies rated per user is about 142, which is much smaller than 10681, the total number of movies. The rating matrix in this situation is extremely sparse. Thus, the mere known ratings are far from enough in predicting the unobserved ratings.

To improve the rating prediction accuracy, the researchers considered both the explicit feedback (such as ratings) and implicit feedbacks (such as some observable behaviors of users, i.e., duration and repetition)~\cite{D.W.Oard}~\cite{X.Kong} in the recommender system~\cite{Xiao:2020,Li:2021,Chen:2021}. Some auxiliary information such as users' reviews~\cite{G.Ling} and the category of items~\cite{Y.Moshfeghi} have also been considered as implicit feedback to improve the recommendation performance. Moreover, Wang et al.~\cite{C.Wang} proposed to combine the probabilistic topic modeling and conventional collaborative filtering to provide an interpretable latent structure between the users and items. The deep learning algorithm has become another direction for the recommender system, since it can be used to obtain the deep representation of auxiliary information and improve the rating prediction accuracy~\cite{M.Fu}~\cite{F.Xia1}~\cite{F.Xia2}. Wang et al.~\cite{H.Wang} proposed a deep learning recommendation model, which adopted the stacked denoising auto-encoder (SDAE) based on probabilistic matrix factorization (PMF)~\cite{Salakhutdinov:2008} to obtain the accurate deep representations. Furthermore, in order to obtain comprehensive features, some research works have utilized the deep learning algorithm in the fields of Natural Language Processing (NLP)~\cite{Y.Kim}~\cite{J.Pennington} and Computer Vision (CV)~\cite{L.Zhao} in the recommender systems. Especially, as for learning word embedding in the NLP field, Lai et al.~\cite{S.Lai} proposed the recurrent convolutional neural network (RCNN), which can capture the entire context information.

Recently, Kim et al.~\cite{D.Kim} proposed a convolutional matrix factorization model (ConvMF), in which the convolutional neural network (CNN) and PMF are integrated to extract the contextual information features of documents and solve the problem that CNN cannot be directly applied to recommendation. Moreover, Chen et al.~\cite{H.Chen} proposed the deformable convolutional matrix factorization (DCNMF), which can capture more contextual information by adding offset layers in convolution layers to further improve the rating prediction accuracy. However, the aforementioned context-aware recommender systems can achieve limited performance improvement, since they just considered the textual information. Actually, the visual information is essential and primary in recognizing the world for the human beings. And the rich visual information of an item can well assist a user to determine whether he is interested in this item or not. For example, a vivid movie poster can help a user to judge whether he likes it or not~\cite{L.Zhao}, since it is difficult to obtain an intuitive and deep impression by just reading the description of the movie.

In this paper, we propose a PMF based movie recommendation scheme called \textit{visual recurrent convolutional matrix factorization} (VRConvMF), which utilizes both the textual and multi-level visual features extracted from the descriptive texts and posters using RCNN and CNN respectively. Moreover, the confidence mechanism considering the user's preference is adopted to optimize the loss function and improve the rating prediction accuracy. To validate the effectiveness of our proposed model, we investigate the impacts of different single-level visual features on the recommendation performance. We validate the performance of the proposed VRConvMF on three different real-world datasets. The experiment results illustrate that the proposed VRConvMF significantly outperforms the existing models.

{\bf The main contributions of this paper} are as follows:

\begin{itemize}

\item We propose the visual recurrent convolutional matrix factorization based recommender scheme called VRConvMF, which can extract the textual features and multi-level visual features to alleviate the sparsity problem of user-to-item rating data in the movie recommender system.

\item We adopt the confidence mechanism in the loss function to improve the rating prediction accuracy.

\item We implement the proposed VRConvMF model and conduct extensive experiments on three real-world datasets to validate its effectiveness.

\end{itemize}

The rest of this paper is organized as follows. Section~\ref{sec2} discusses related works of recommendation. Section~\ref{sec3} introduces matrix factorization method and briefly describes word embedding representation, convolutional neural network for text and confidence mechanism. Section~\ref{sec4} introduces the proposed movie recommender system VRConvMF in details. Section~\ref{sec5} conducts the performance evaluation. Section~\ref{sec6} summarizes this paper and outlines the future work.

\section{Related Work}\label{sec2}

In this section, we briefly discuss the collaborative filtering algorithm and the context-aware recommendations based on neural networks.

 \subsection{Collaborative Filtering}
 The collaborative filtering (CF) algorithm is to obtain the required recommendations by collaborative processing of the ratings given by a large number of users. Specifically, matrix factorization is one of the most commonly used methods in CF, which will be discussed in details in~\ref{sec3.1}. Since the observed ratings in user-to-item rating matrix are closely related to users' preference for items, the basic idea of CF is to estimate missing ratings from observed ratings. Furthermore, CF can be roughly divided into two types, one is the user-based CF (UserCF), and the other is the item-based CF (ItemCF). Specifically, UserCF makes use of user ratings, which are similar to target user~$i$'s preference, to predict user~$i$'s recommendations. The similarity function is used to find similar users by calculating each line of the rating matrix. While ItemCF is used to calculate the item set $S$, which is similar to item~$j$. Then the ItemCF can predict whether user~$i$ appreciates item~$j$ or not based on user~$i$'s ratings to other items in~$S$. It is easy to implement the traditional CF and the recommendations generated by CF are highly interpretable. However, CF is not suitable for the sparse rating matrix since it is difficult to find similar users or items. Therefore, alleviating the sparsity of the rating matrix is significant to improve the performance of collaborative filtering based recommendation.

  \subsection{Context-aware Recommendations}
Recently, deep learning methods based on neural networks (such as recurrent neural networks (RNN), CNN and their derivative networks) have been widely applied to recommender systems~\cite{D.Kim}~\cite{H.Chen}~\cite{P.Covington}. Specifically, RNN are often used to process textual information, while CNN are more commonly used to process image information. Recommender systems can obtain the deep-level representation of various data types by using deep learning methods~\cite{S.Duan}. These deep-level representations are able to obtain the essential characteristics of items and users~\cite{Z.Huang}.

Context-aware recommendations utilize the contextual information to improve the rating prediction accuracy of recommender systems.
  Generally speaking, contextual information is extracted from attributes of items and user comments. A lot of additional information (such as time stamps~\cite{Y.Koren1}, locations~\cite{H.Stormer} and social networks~\cite{A.Q.Macedo}) can be used as contextual information for recommender systems. These contextual information is interdependent or independent. Many works considered contextual information such as social networks~\cite{S.Purushotham}, user comment texts~\cite{J.McAuley} and description of items~\cite{H.Wang} to improve the recommendation performance so far. Accordingly, as for recommender systems with specific tasks, the contextual information is different. For instance, music recommendation utilized some specific information, such as acoustic features to improve the performance of recommendation with sequence of songs~\cite{H.Negar}~\cite{R.Cheng}. Furthermore, contextual information such as singer information, date of publish and the style of the song are applied in music recommendation to solve the cold-start problem~\cite{S.Oramas}. As for movie recommendation, contextual information contains a wealth of attributes such as actor, director, genre, poster, trailer and descriptive text. Zhao et al.~\cite{L.Zhao} presented movie recommendation model using visual features extracted from picture data. Moreover, Fan et al.~\cite{Y.Fan} extracted visual features from trailers~\cite{Vinyals:2015} directly using Youtube-8M dataset~\cite{S.A.-E.-H.}. Compared to all of the above forms of contexts, there are relatively few of works considering the fusion of multiple features (such as textual features and visual features) to recommender systems.

\section{Preliminary}\label{sec3}

\subsection{Matrix Factorization}\label{sec3.1}
Matrix factorization (MF) is one of the collaborative filtering algorithms in the application of recommendation~\cite{G.Adomavicius}~\cite{Y.Koren}. The purpose of MF is to predict missing ratings by using observed ratings. In the traditional MF, it is supposed that there are $m$ users, $n$ items and a user-to-item rating matrix $\mathbf{R}\in\mathbb{R}^{m\times n}$. Each row represents a user and each column represents an item. Each element $r_{ij}$ in $\mathbf{R}$ represents the rating of user $i$ ($1$$\leq$$i$$\leq$$m$) on movie $j$ ($1$$\leq$$j$$\leq$$n$). Typically, ratings are divided into several levels (i.e., from 1 to 5), where higher values mean stronger preference. If user $i$ did not rate movie $j$, let $r_{ij} = 0$. The $k-dimensional$ latent models of users and latent models of items are represented as the user-factor matrix $\mathbf{U}\in\mathbb{R}^{k\times m}$ and the item-factor matrix $\mathbf{V}\in\mathbb{R}^{k\times n}$ respectively. Accordingly, the $i^{th}$ user and $j^{th}$ item are denoted as $u_{i}$ and $v_{j}$ respectively. Each rating $r_{ij}$ of user $i$ on item $j$ can be approximated by inner product of corresponding latent models of user $i$ and item $j$: $r_{ij}\approx\widehat{r_{ij}}= u_{i}^T v_{j}$. Generally, training latent models tends to minimize the sum of squared errors between the actual rating $r_{ij}$ and the predictive rating $\widehat{r_{ij}}$, and $L_2$ regularization term is applied to avoid the over-fitting problem. The loss function $\mathscr{L}$ can be expressed as:
\begin{equation}
\resizebox{.91\linewidth}{!}{$
\displaystyle
\mathscr{L}=\sum_{i=1}^m\sum_{j=1}^n I_{ij}\Big(r_{ij} - u_i^T{v_j}\Big)^2+ \lambda_{u}\sum_{i=1}^m\Big|\Big|u_{i}\Big|\Big|^2 + \lambda_{v}\sum_{j=1}^n\Big|\Big|v_{j}\Big|\Big|^2, \label{eq1}$}
\end{equation}
where $I_{ij}$ is the index function which is shown as follows.
\begin{displaymath}
I_{ij}=\left\{\begin{array}{l@{\ \
\ }l}
1,&\textrm{if user $i$ rated item $j$,}\\
0,&\textrm{elsewhere.}\\
\end{array}\right.
\end{displaymath}

\subsection{Word Embedding Representation}
Word embedding is the first step in deep learning methods for natural language processing. Global-Vector (GloVe)~\cite{J.Pennington} is one of the models that can convert natural language into machine language. GloVe model can learn textual information according to the full text, and can obtain word embedding vectors, which are represented by dense numeric matrices. The obtained word embedding vectors are able to capture the semantic properties between different words (such as similarity and analogy). The semantic similarity between two words can be measured via calculation based on the word embedding vectors (such as Euclidean distance or cosine similarity). GloVe model can be approximately divided into three parts. The first part is to build a co-occurrence matrix $\mathbf{Co}$ according to the corpus, each element $co_{ij}$ in $\mathbf{Co}$ represents the number of times that word $i$ and its context word $j$ coexist in a context window with a specific size. Instead of using natural numbers, the decreasing weighting function $decay = 1/d$ is applied to count the number of word co-occurrences in GloVe model. Specifically, $d$ denotes the distance between the context windows of two words. Obviously, in the total count, the further the distance between two words is, the smaller the corresponding value of $decay$ will be. The second part is to build the approximate relationship of word embedding vectors and co-occurrence matrix as Eq.~(\ref{eq2}):
\begin{equation}
w_i^T\widetilde{w}_j + b_i + \widetilde{b}_j \approx log\Big(co_{ij}\Big),
 \label{eq2}
\end{equation}
where $w_i^T$ and $\widetilde{w}_j$ represent the word embedding vectors, $b_i$ and $\widetilde{b}_j$ denote the bias terms of the word embedding vectors $w_i^T$ and $\widetilde{w}_j$ respectively. While the third part is to build the loss function as Eq.~(\ref{eq99}):
\begin{equation}
J = \sum_{i,j=1}^V f(co_{ij})\Big(w_i^T\widetilde{w}_j + b_i + \widetilde{b}_j - log(co_{ij})\Big)^2,
 \label{eq99}
\end{equation}
where,
\begin{displaymath}
f(x)=\left\{\begin{array}{l@{\ \
\ }l}
\Big(x/x_{max}\Big)^{\beta},&\textrm{if $x$ $<$ $x_{max}$,}\\
1,&\textrm{elsewhere.}\\
\end{array}\right.
\end{displaymath}

In the GloVe model, the adagrad method is utilized to optimize the parameters in loss function $J$ to obtain the word vector $w$ that is closest to the non-zero elements in the co-occurrence matrix $\mathbf{Co}$. The application of the co-occurrence matrix can easily obtain the full-text contextual information by changing different context window sizes.
\begin{figure}
  \includegraphics[width=3.7in]{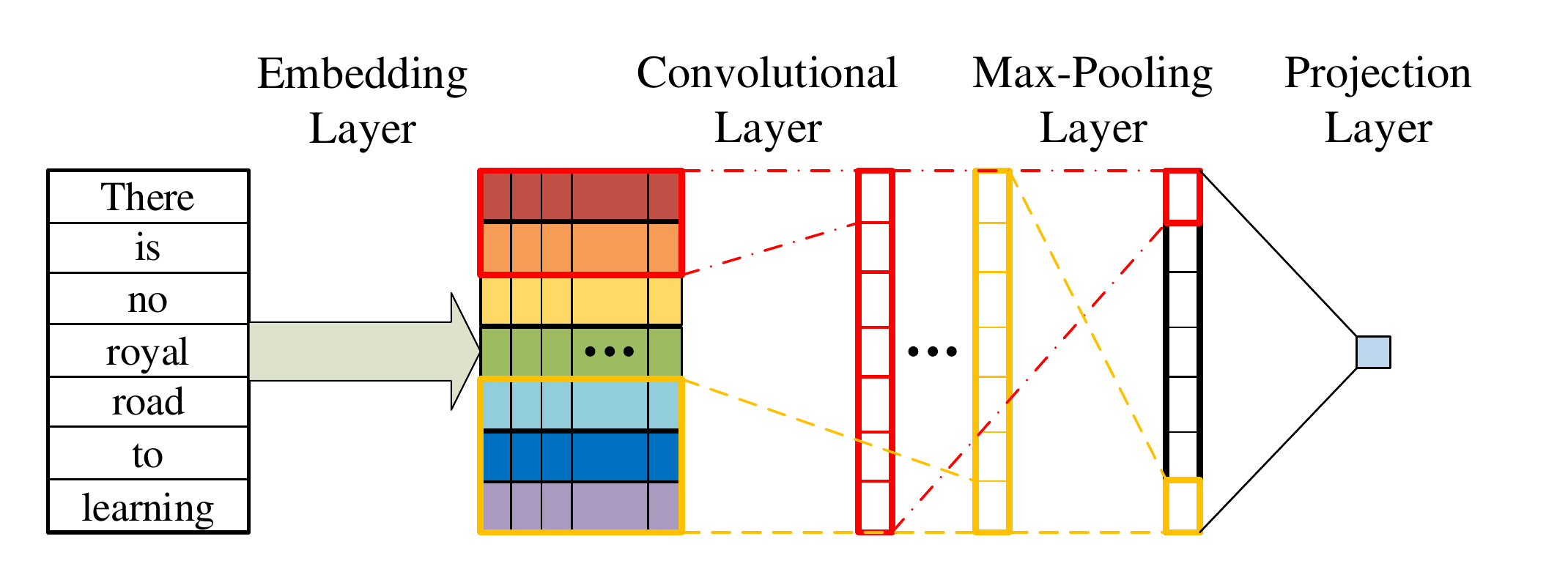}
  \caption{Convolutional neural network for text.}
  \label{figure1}
\end{figure}

\subsection{Convolutional Neural Network for Text}
\subsubsection{Embedding Layer}
The embedding layer is the first step to input natural language into mathematical models. We refer to each text as a sequence of words and transform all the sequences of words into a dense numeric matrix. In detail, all of the words are converted to vectors by pre-trained word embedding model (such as $Word2Vec$~\cite{T.Mikolov} and $GloVe$), then cascade all of the word vectors to represent textual documents mathematically. The descriptive texts or users' reviews of movie $j$ are expressed as:
\begin{equation}
X_j={\Big[...{w_{i - 1}}\oplus{w_i}\oplus{w_{i + 1}}...\Big]^T_{p \times q}}, \label{eq13}
\end{equation}
where $p$ is the dimension of each word embedding vector, $q$ is the length of descriptive texts of movies, and $\oplus$ represents cascade operation. Specifically, we use attributes of movies for MovieLens dataset and utilize users' reviews as descriptive texts for AIV dataset.

\subsubsection{Convolutional Layer}

A conventional convolution operation in sentence classification involves filters with the $j^{th}$ shared weight $W_j =\mathbb{R}^{h\times p}$, where $h$ is the convolutional window size. Specifically, a CNN with a small convolutional window size results in losing some far distance contextual information, while a large convolutional window size can capture more contextual information, however, this situation will lead to data sparsity. A textual feature $c_{j}=g\big(W_{j}\bullet w_{i:i+h-1}+b_{j}\big)$ is obtained by using convolution operation, where $\bullet$ represents convolution operator, $i$ represents the $i^{th}$ word embedding, $b_{j}$ is the bias term for $W_j$, and $g\big(\cdot\big)$ is a non-linear activation function such as sigmoid, hyperbolic tangent and rectified linear unit ReLU. Then, all of the filters with different window sizes are applied to traverse each word vector of the entire text $\Big\{w_{1:h},w_{2:h+1},...,w_{q-h+1:q}\Big\}$. A contextual feature vector of a textual document is given by:
\begin{equation}
\mathbf{C}_{t}=\Big[c_{1},c_{2},...,c_{q-h+1}\Big], \label{eq14}
\end{equation}
where $\mathbf{C}_t$ refers to the textual feature of the $t^{th}$ document extracted by convolution operator. Moreover, the multiple shared weights are used to capture the multiple textual features.

\subsubsection{Pooling Layer}

The pooling layer is applied to further extract the textual features of the convolutional layer in Eq.~(\ref{eq14}). Max pooling operation is applied over the feature map to reduce the dimension of the textual feature vector to a fixed length $n_c$ as follows:
\begin{equation}
\mathbf{D}=\Big[max(C_{1}),max(C_{2}),...,max(C_{n_{c}})\Big], \label{eq15}
\end{equation}

\subsubsection{Projection Layer}

$\mathbf{D}$ is projected in Eq.~(\ref{eq15}) on a $k-dimensional$ space of item latent models for VConvMF model. The final textual feature is obtained by using conventional nonlinear projection as follows:
\begin{equation}
\mathbf{\Phi}=tanh\bigg\{W_{fc2}\Big[tanh(W_{fc1}\mathbf{D}+b_{fc1})\Big] +b_{fc2}\bigg\}, \label{eq16}\
\end{equation}
where $\mathbf{W}_{fc1}\in \mathbb{R}^{x\times n_{c}}$ and $\mathbf{W}_{fc2}\in \mathbb{R}^{k\times x}$ are the weights of fully connected layers, and $x$ is the number of hidden layer nodes for fully connected layer.
The framework of convolutional neural network for text is shown in Fig.~\ref{figure1}.

\subsection{Confidence Mechanism}
In this section, the confidence levels of expressing preferences between different users' ratings are considered. Specifically, extreme ratings (such as 1 and 5) are more reliable than moderate ratings (such as 2, 3 and 4). Therefore, the confidence value $c_{ij}$ of each rating is adopted to revise the loss function. The confidence mechanism is used to assign higher values for more effective ratings. The specific confidence factor expression is as follows:
\begin{equation}
c_{ij}=1+\alpha\cdot f\Big(r_{ij}-\frac{r_{max}}{2}\Big), \label{eq12}
\end{equation}
where $\alpha$ is a hyper-parameter to control the value of the confidence factor $c_{ij}$, $f(\cdot)$ is a distance function (such as the absolute function or square function), and $r_{max}$ is the maximum of ratings. Obviously, $r_{max}=5$.

\section{The Movie Recommendations}\label{sec4}

\subsection{Textual Feature Extraction in Movies}\label{sec4.1}
 The RCNN model, which is designed for text classification~\cite{S.Lai}, is adopted for the textual feature extraction in the proposed VRConvMF. The extracted features are used as a part of the mean of gaussian distribution in the item latent models. In order to obtain more complete contextual information of the text, the recurrent structures are integrated into convolutional layers to further improve the quality of word representations. In this paper we propose the recurrent convolutional matrix factorization called \textit{RConvMF}, which can take advantages of both the RCNN and PMF. In RCNN models, the word representations combine the word and its context together to understand the word more comprehensively. It is supposed that $ct_l\big(w_i\big)$ represents the left context of word $w_i$ and $ct_r\big(w_i\big)$ represents the right context of word $w_i$. The left and right contexts of word $w_i$ are defined as:
\begin{equation}
ct_l(w_i)= f\bigg(\mathbf{W}^{(l)}ct_l\Big(w_{i-1}\Big)+\mathbf{W}^{(sl)}e\Big(w_{i-1}\Big)\bigg), \label{eq3}
\end{equation}
\vspace{-2mm}
\begin{equation}
ct_r(w_i) = f\bigg(\mathbf{W}^{(r)}ct_r\Big(w_{i+1}\Big)+\mathbf{W}^{(sr)}e\Big(w_{i+1}\Big)\bigg), \label{eq4}
\end{equation}
where $e\big(w_i\big)$ denotes the word embedding of word $w_i$, $\mathbf{W}^{(l)}$ and $\mathbf{W}^{(r)}$ denote the matrices, which combine all the left and the right context hidden layers respectively, $\mathbf{W}^{(sl)}$ and $\mathbf{W}^{(sr)}$ denote the matrices, which are used to combine the semantic of current word with the left and right contexts of next word respectively, $f\big(\cdot\big)$ is a nonlinear activation function, i.e., ReLU. Under this model, the word representation $x_i$ of the word $w_i$ is denoted as follows:
\begin{equation}
x_i = \bigg[ct_l\Big(w_i\Big), e\Big(w_i\Big), ct_r\Big(w_i\Big)\bigg].  \label{eq5}
\end{equation}

Obviously, $ct_l\big(w_i\big)$ and $ct_r\big(w_i\big)$ represent all the left and right contextual information of word $w_i$ respectively, and different context window sizes are used to capture different contextual information in order to investigate the performance of RCNN model. For instance, the word representation of $w_i$ is represented by $\Big[x\big(w_{i-2}\big); x\big(w_{i-1}\big); x\big(w_i\big); x\big(w_{i+1}\big); x\big(w_{i+2}\big)\Big]$ when the context window size is 5. Furthermore, an activation function, i.e., $tanh$, is applied to transform $x_i$ into $x_i^{(2)}$ as follows:
\begin{equation}
x_i^{(2)} = tanh\Big(\mathbf{W}^{(2)}x_i + b^{(2)}\Big). \label{eq6}
\end{equation}

Max-pooling layer is applied after all word representations are obtained, and max-pooling operation is able to capture the most significant factors of the whole text. The further textual features can be obtained as follows:
\begin{equation}
x^{(3)} = \max\limits_{i\in \{1,2,\cdots,n\}} x_i^{(2)}. \label{eq7}
\end{equation}

Then, the conventional linear function is applied in neural networks to project the output from pooling layer on $k-dimensional$ space of item latent models. The final textual feature of descriptive text of movie $j$ is obtained as follows:
\begin{equation}
v_{j} = \mathbf{W}^{(4)}x^{(3)}+b^{(4)}. \label{eq8}
\end{equation}
For textual feature extraction, the RCNN architecture can be expressed as $\varphi_{j}=rcnn\big(\mathbf{W},X_{j}\big)$,
where $\mathbf{W}$ denotes all of the weights and biases, and $\varphi_{j}$ denotes the textual latent vector of movie $j$.

\begin{figure*}

  \includegraphics[width=7in,height=2.6in]{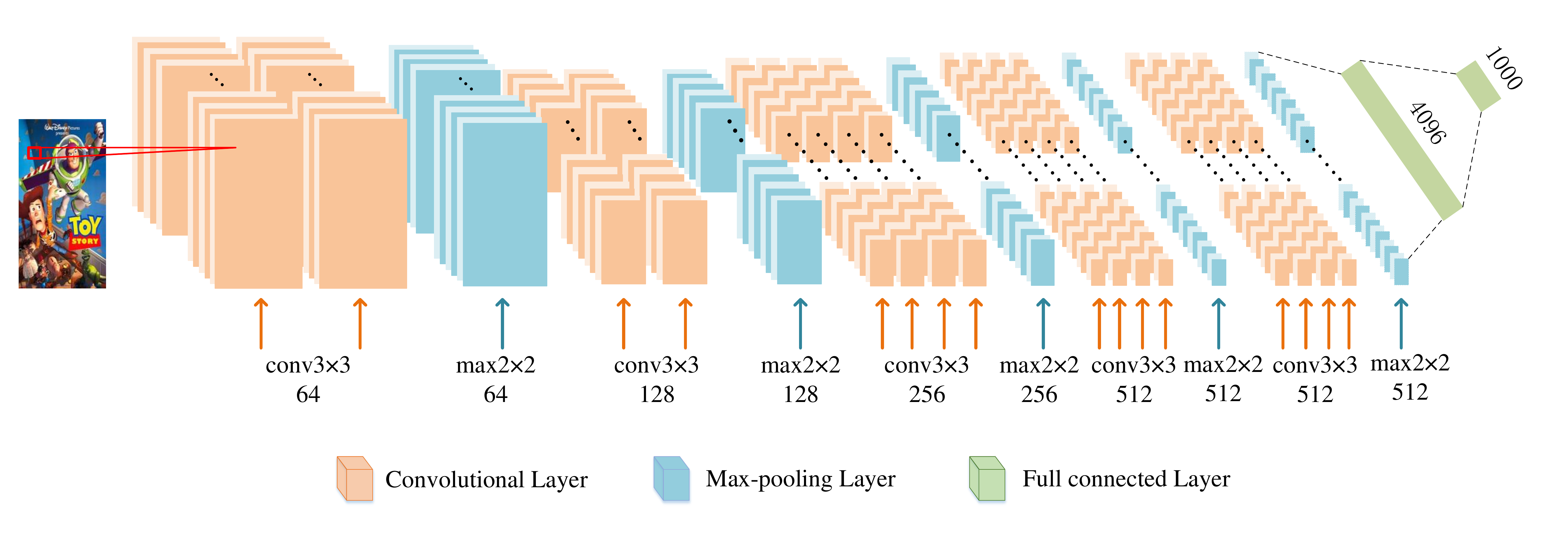}
  \caption{The architecture of VGG19.}
  \label{figure2}

\end{figure*}
\subsection{VGG of Visual Feature Extraction}\label{sec4.2}
 VGG is a kind of CNN with a deep structure designed by \textit{Visual Geometry Group}. VGG can fully extract the features of each picture~\cite{K.Simonyan}. Generally speaking, high-level visual features focus more on semantic information while less on details. Low-level visual features contain more detailed information while they are related to background confusion and semantic ambiguity~\cite{W.Yu}. Specifically, the VGG19 model is utilized to extract visual features, which contains 19 weight layers, i.e., 16 convolutional layers and 3 fully connected layers. Moreover, the size of all of the convolution kernels is $3\times 3$, which can get the pixel values of adjacent positions with the minimum size. In addition, all of the max-pooling operators are performed with $2\times 2$ pixel windows. All of the hidden layers in VGG19 are equipped with ReLU, a non-linearity activation function. The specific architecture of VGG19 is shown in Fig.~\ref{figure2}. We can extract visual features at different depth levels from several specific layers. Different single-level visual features are obtained from different depth pooling layers respectively. The feature cross technology is applied to cascading the multi-level information. Specifically, different depth visual features of each image are cascaded as the multi-level visual feature. For visual feature extraction, the VGG19 model is summarized as follows: $\phi_{j}=vgg\big(\mathbf{W}',Y_{j}\big),$
where $Y_{j}$ denotes the image of item $j$ (including posters and still frames), and $\phi_{j}$ denotes visual features of movie $j$.

\subsection{Probabilistic Model of RConvMF}
The probabilistic model of RConvMF can take fully advantage of both item description documents and ratings by bridging RCNN and PMF. The conditional distribution over observed ratings is as follows:
\begin{equation}
p\Big(\mathbf{R}|\mathbf{U},\mathbf{V},{\sigma ^2}\Big)=\mathop \prod \limits_{i = 1}^m \mathop \prod \limits_{j = 1}^n {\bigg[ N\Big({r_{ij}}|u_i^T{v_j},{\sigma ^2}\Big)\bigg]^{{I_{ij}}}},\label{eq9}
\end{equation}
where $N\big(x|\mu ,{\sigma ^2}\big)$ is the probability density function of the gaussian distribution with mean $\mu$ and variance $\sigma ^2$, and $I_{ij}$ is the index function, which has been mentioned in Section~\ref{sec3.1}. Then, zero-mean spherical gaussian prior with variance $\sigma _U^2$ is placed on user feature vectors:
\begin{equation}
p\Big(\mathbf{U}|\sigma _\mathbf{U}^2\Big)=\mathop \prod \limits_{i = 1}^m N\Big({u_i}|0,\sigma _\mathbf{U}^2I\Big). \label{eq10}
\end{equation}

Accordingly, in conventional PMF, item feature vectors with zero-mean spherical gaussian priors are represented by $p(\mathbf{V}|\sigma _\mathbf{V}^2)=\mathop \prod \limits_{j = 1}^n N\big({v_j}|0,\sigma _\mathbf{V}^2I\big)$. Moreover, in probabilistic model of RConvMF, the key idea of connecting the PMF and RCNN is to use a document latent vector obtained from the RCNN model as the mean of gaussian distribution, and gaussian noise of the item as the variance of gaussian distribution. Accordingly, $v_{j}$ is changed into ${v_j}=rcnn\big(\mathbf{W},{X_j}\big)+{\varepsilon _j}$, where $\varepsilon _{j}\sim N\big(0,\sigma _\mathbf{V}^2I\big)$. The conditional distribution over item latent models is given by:
\begin{equation}
p\Big(\mathbf{V}|\mathbf{W},\mathbf{X},\sigma_\mathbf{V}^2\Big)=\mathop \prod \limits_{j=1}^n N\bigg({v_j}|rcnn\Big(\mathbf{W},{X_j}\Big),\sigma _\mathbf{V}^2I\bigg), \label{eq11}
\end{equation}
where $\mathbf{X}$ is the set of descriptive texts of items (such as description of movies and user reviews), and $X_{j}$ represents the description of movie $j$. For the weight $w$ in $\mathbf{W}$, which exists in RCNN, we place zero-mean spherical gaussian prior: $p\big(\mathbf{W}|\sigma _\mathbf{W}^2\big) = \mathop \prod \limits_k N\big({w_k}|0,\sigma _\mathbf{W}^2\big)$.
\vspace{-7mm}

\subsection{VRConvMF Model}
In this Section, the proposed model will be further explained. In Sections \ref{sec4.1} and \ref{sec4.2}, the extractions of the textual features and visual features are introduced respectively. Then the above two features are merged by using the following method and integrate recurrent neural network and convolutional network as textual feature extraction. Then we separately cascade the textual features and the corresponding visual features of movies as the comprehensive features and project the comprehensive features through projection layer to specific dimensions.

\begin{figure*}
  \centering
  \includegraphics[width=6.8in,height=3.6in]{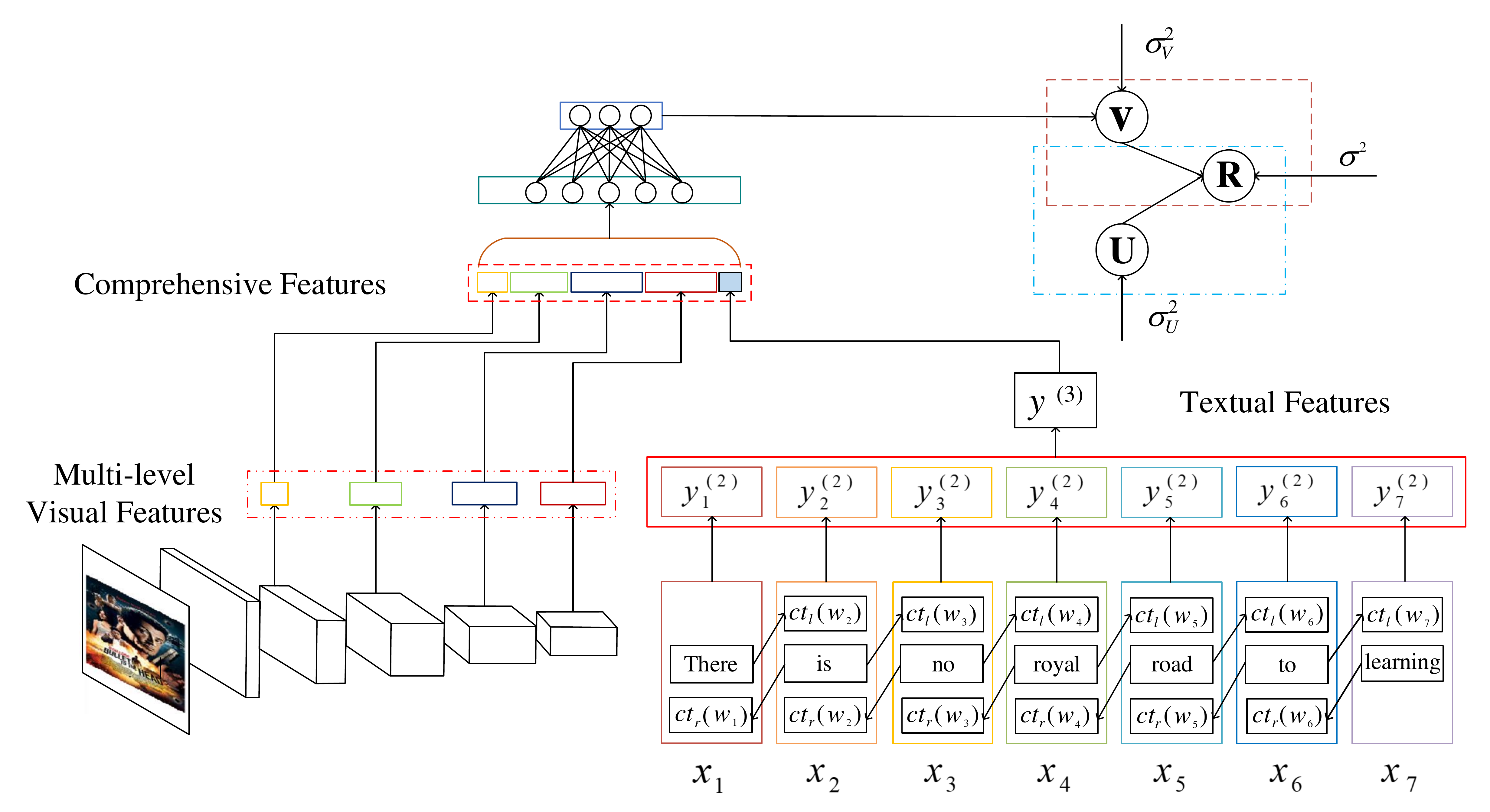}
  \caption{The architecture of VRConvMF model.}
  \label{figure3}
\end{figure*}

Accordingly, as for conditional distribution over item feature vectors, we intensify them on the basis of traditional PMF by:
\begin{equation}
p\Big(\mathbf{V}|\mathbf{W'},\mathbf{X},\mathbf{Y},\sigma_V^2\Big)=\mathop\prod\limits_{j=1}^n N\Big(v_{j}|\mu_{j},\sigma_\mathbf{V}^2 I\Big),\label{eq17}
\end{equation}
where $\mu_{j}=dense\Big(cascade\big(\varphi_{j},\phi_{j}\big)\Big)$, $\mathbf{W'}$ represents the weights and biases of VGG, RCNN and dense layer, and $\mathbf{Y}$ is the set of images of movies. Then, the item latent factor $v_j$ can be expressed as follows:
\begin{equation}
v_{j}=dense\bigg(cascade\Big(\varphi_j,\phi_j\Big)\bigg)+\varepsilon_j, \label{eq18}
\end{equation}
where $\varepsilon_j\sim N\big(0,\sigma_\mathbf{V}^2 I\big),$ and the $cascade$ refers to concatenating the textual features obtained from the RCNN model and the visual features obtained from the VGG network. Then, the features from two different vector spaces can be fused as the comprehensive features.
Finally, the obtained $v_{j}$ is adopted in PMF. The framework of the proposed VRConvMF is shown in Fig.~\ref{figure3}.
We adopt the Maximum A Posteriori (MAP) estimation to optimize the variables, weights and biases as follows:

\begin{equation}
\begin{split}
&\max\limits_{\mathbf{U},\mathbf{V},\mathbf{W'}}p\bigg(\mathbf{U},\mathbf{V},\mathbf{W'}\Big|\mathbf{R},\mathbf{X},\mathbf{Y},\sigma^2,\sigma_\mathbf{U}^2,\sigma_\mathbf{V}^2,\sigma_{\mathbf{W}'}^2\bigg)\\
&=\max\limits_{\mathbf{U},\mathbf{V},\mathbf{W'}}\bigg[p\Big(\mathbf{R}\Big|\mathbf{U},\mathbf{V},\sigma^2\Big)p\Big(\mathbf{U}\Big|\sigma_\mathbf{U}^2\Big)p\Big(\mathbf{V}\Big|\mathbf{W'},\mathbf{X},\mathbf{Y},\sigma_\mathbf{V}^2\Big)p\Big(\mathbf{W'}\Big|\sigma_{\mathbf{W'}}^2\Big)\bigg].\\\label{eq19}
\end{split}
\end{equation}

The confidence mechanism is adopted in loss function, and the negative logarithm of the posterior distribution over the user features and movie features in Eq.~(\ref{eq19}) is given by:
\begin{equation}
\begin{split}
\mathcal{L}(\mathbf{U},\mathbf{V},\mathbf{W'})&=\sum\limits_{i}^m\sum\limits_{j}^n\frac{{{c_{ij}}}}{2}\bigg[I_{ij}\Big(r_{ij}-u_i^T{v_j}\Big)^2+\lambda_{\mathbf{U}}\sum_{i=1}^m\Big|\Big|u_{i}\Big|\Big|_2\\
                      &+\lambda_{\mathbf{V}}\sum_{j=1}^n\Big|\Big|v_{j}-\mu_{j}\Big|\Big|_2+\lambda_{\mathbf{W'}}\sum_k^{w'_k}\Big|\Big|w'_k\Big|\Big|_2\bigg],\\\label{eq20}
\end{split}
\end{equation}
where ${\lambda _\mathbf{U}} = \sigma^2 / \sigma_\mathbf{U}^2,$ ${\lambda _\mathbf{V}} = \sigma^2 / \sigma_\mathbf{V}^2,$ ${\lambda _{\mathbf{W}'}} = \sigma^2 / \sigma_{\mathbf{W}'}^2.$

For the optimization process, the coordinate descent method is adopted to iteratively optimize the latent variable while fixing remaining variables. Specifically, coordinate descent method performs a one-dimensional search along the coordinate direction at the current point to find the minimum of loss function. In addition, the optimal solution of $\mathbf{U}$ and $\mathbf{V}$ can be calculated by the loss function $\mathcal{L}$ with respect to $u_{i}$ and $v_j$ respectively as follows:
\begin{equation}
{u_i} \leftarrow {\Big(\mathbf{V}{I_i}{\mathbf{V}^T} + {\lambda _\mathbf{U}}{I_K}\Big)^{ - 1}}\mathbf{V}{R_i},\label{eq21}
\end{equation}
\begin{equation}
{v_j} \leftarrow {\Big(\mathbf{U}{I_j}{\mathbf{U}^T} + {\lambda _\mathbf{V}}{I_K}\Big)^{ - 1}}\Big(\mathbf{U}{R_j} + {\lambda_V}{\mu_j}\Big),\label{eq22}
\end{equation}
where $I_{i}$ is a diagonal matrix with $I_{ij}$, $j=1,...n$ as its diagonal elements. And $R_{i}$ is a vector for user $i$ with $\big(r_{ij}\big)_{j=1}^n$. For movie $j$, $I_{j}$ and $R_{j}$ are similarly defined as $I_{i}$ and $R_{i}$ respectively. The back-propagation algorithm is applied to optimize $\mathbf{W}'$. In the whole optimization process, the unobserved rating of user $i$ on movie $j$ can be predicted as: $\widehat{r_{ij}}\approx E\Big[r_{ij}\Big|u_i^T{v_j},\sigma^2\Big]=u_i^T{v_j}=u_i^T\Big(\mu_j + \varepsilon_j\Big).$

\section{Experiments}\label{sec5}

\subsection{Experimental Settings}

\subsubsection{\textbf{Datasets}}

We validate the proposed VRConvMF on three different real-world datasets including MovieLens-1m, MovieLens-10m and Amazon Instant Video. In the following parts of this paper, the three datasets are abbreviated as ML-1m, ML-10m and AIV respectively.
\begin{itemize}
\item $\textbf{ML-1m}$: It is a movie rating dataset which includes the user ratings (each of which is in terms of a value from 1 to 5), metadata of movies (Movie ID, title and genres) and user attribute information (User ID, gender, age and occupations), etc. Specifically, it contains 6040 users, 3883 movies and 1000209 ratings. It is called 1m because the dataset contains 1M rating data.

\item $\textbf{ML-10m}$: Similar to ML-1m, it is also a dataset which is provided by MovieLens website. The dataset contains 71567 users, 10681 movies and 10000054 ratings specifically. And the movies can be classified into 19 categories according to our statistics.

\item $\textbf{AIV}$: The Amazon rating dataset contains the user ratings and the metadata of movies from Amazon website. It contains 29757 users, 15149 items and 135188 ratings. According to statistics, there are 21 categories for all the movies.
\end{itemize}
\begin{table}
\caption{Data statistics on three real-world datasets.}\label{table1}
\begin{spacing}{1}
\centering
\label{table}
\begin{tabular}{c c c c c}
\toprule[1.5pt]
{Datasets}&Users&Movies&Ratings&Sparsity\\
\bottomrule[1.5pt]
ML-1m&6040&3883&1000209&95.73\%\\
\hline
ML-10m&71567&10681&10000054&98.69\%\\
\hline
AIV&29757&15149&135188&99.97\%\\
\toprule[1.5pt]
\end{tabular}
\end{spacing}
\end{table}

 Specifically, ML-1m and ML-10m were obtained from $MovieLens$\footnote{http://files.grouplens.org/datasets/movielens/}, and the AIV was obtained from $Amazon$ $website$ $of$ $video$\footnote{http://jmcauley.ucsd.edu/data/amazon/}. Due to the lack of corresponding auxiliary information, specifically, we utilize web crawler technique to crawl movie description documents for MovieLens datasets, users' reviews for AIV dataset, and movie posters or still frames on $IMDb$\footnote{http://www.imdb.com/}, $Netflix$\footnote{http://www.netflix.com/} and
$Douban$\footnote{https://movie.douban.com/} websites for all of the datasets. Before feeding the proposed VRConvMF, auxiliary information is preprocessed for all of the datasets as follows: 1) for the movies without posters, choose the still frames which are extracted from trailers; 2) remove the movies without any text or image information. The sparsity of the three datasets is shown in TABLE~\ref{table}, which illustrates that the sparsity problem of each dataset is extremely serious.

For the textual feature extraction, similar to~\cite{Y.Kim}, we preprocess descriptive texts of movies for all of the datasets as follows: 1) set maximum length of raw text to 400; 2) remove stop words; 3) select top 6000 distinct words as a vocabulary and remove the other words from raw documents. For the visual feature extraction: 1) resize all of the RGB images with the size of $\big(182\times268\times3\big)$; 2) utilize $min-max$ $scaling$ $normalization$ to get feature maps; 3) perform $Histogram$ $Equalization$ to enhance the quality of posters since a few of movies were released in the early time, the posters of which are not sharp enough.
\renewcommand\arraystretch{1.3}

\subsubsection{\textbf{Model Settings}}
The dimension of user latent vectors and item latent vectors \big($\mathbf{U}$ and $\mathbf{V}$\big) are both set to 50 (as reported in~\cite{H.Wang}) and we initialize $\mathbf{U}$ and $\mathbf{V}$ randomly in the range of \big(0,1\big). We implement the proposed VConvMF and VRConvMF by using Python and Keras library with Theano backend. The hardware is $Nvidia 1080Ti$ $GPU$ on Linux Ubuntu platform. The dropout method is used to avoid over-fitting and the dropout rate is set to 0.3 during the training of the RCNN and CNN models.

 The specific settings and implementations of models are as follows:
 \begin{itemize}
\item $\textbf{VConvMF}$:
    1) for the pre-trained word embedding model, we use $GloVe$, and initialize word vectors randomly with the dimension size of 200, which will be trained through the optimization process; 2) various convolutional window sizes \big(3, 4, 5\big) with 100 feature maps are used to obtain different contextual information; 3) mini-batch size is set to 128.
\item $\textbf{VRConvMF}$:
    1) the pre-trained $GloVe$ model is also used as our word embedding model to speed up the training process, 2) the size of word embedding and context window, i.e., $\big|e\big|$ and $\big|c\big|$, and $\beta$ are set to 200, 50 and 0.75 respectively, 3) the hidden layer size $h$ is set to 100.
 \end{itemize}

\begin{figure}[htbp]
\centering

\subfigure{
\begin{minipage}[t]{0.49\linewidth}
\centering
\includegraphics[width=1.7in]{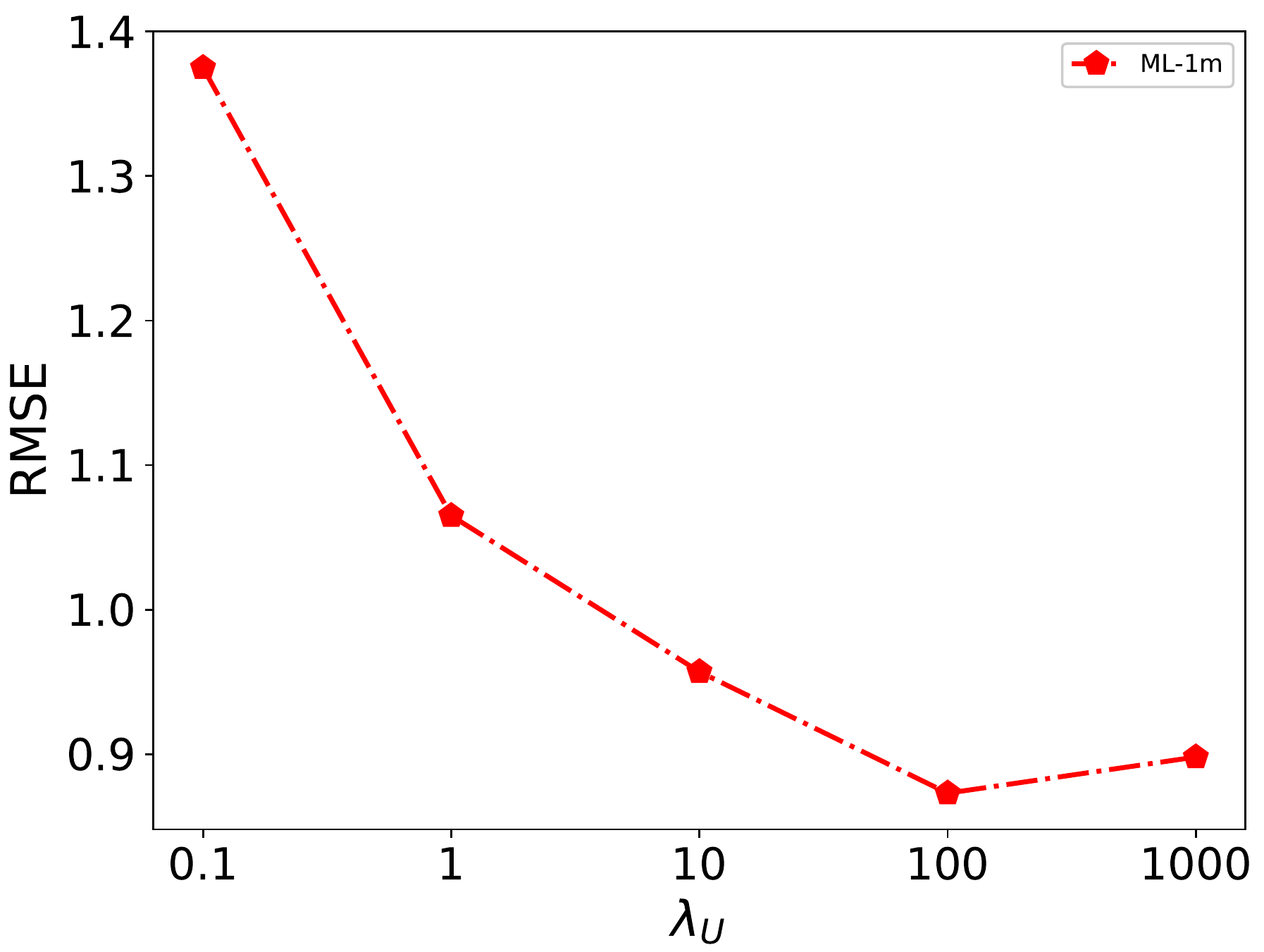}
\end{minipage}%
}%
\subfigure{
\begin{minipage}[t]{0.49\linewidth}
\centering
\includegraphics[width=1.7in]{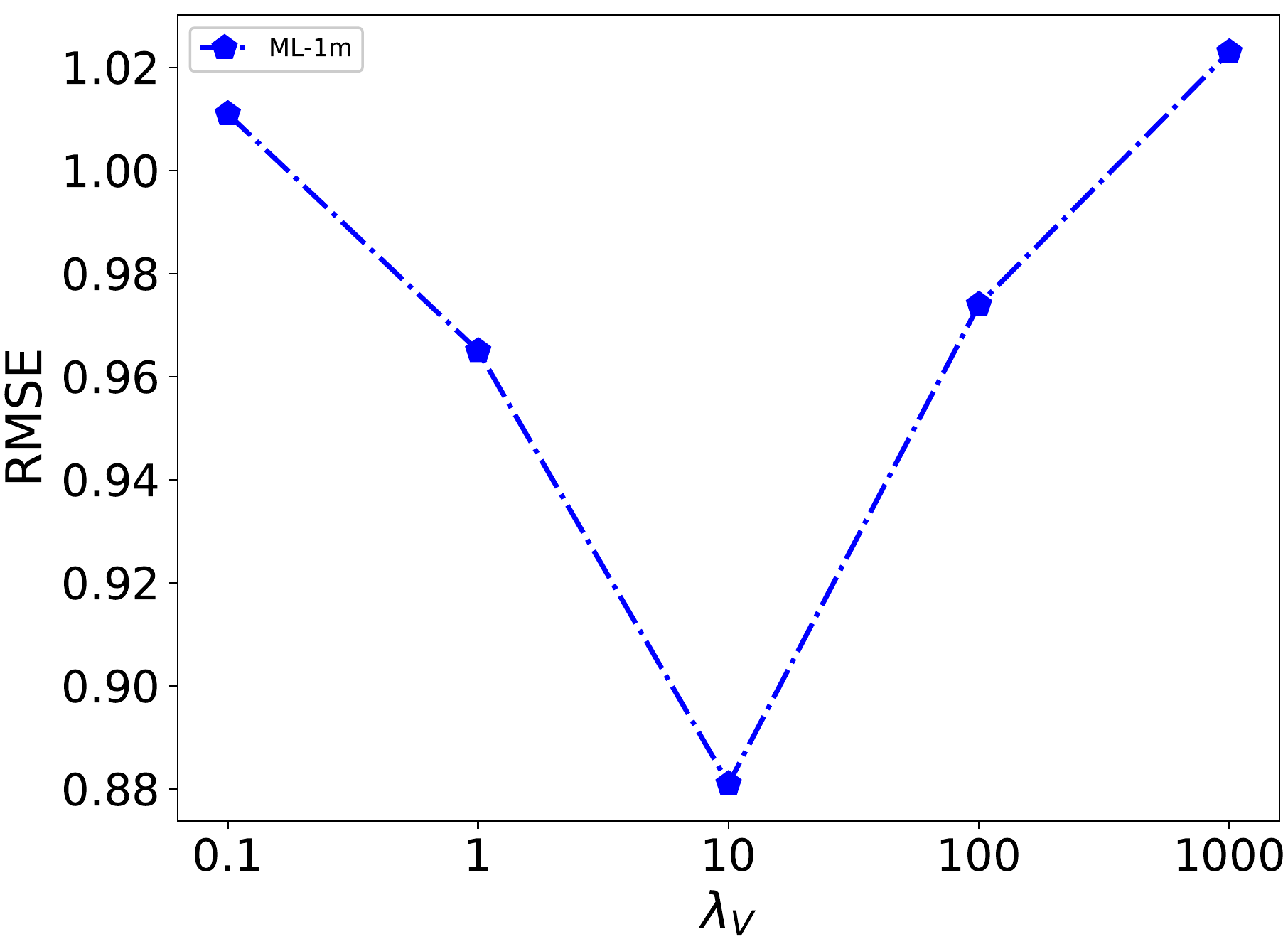}
\end{minipage}%
}%

\subfigure{
\begin{minipage}[t]{0.48\linewidth}
\centering
\includegraphics[width=1.7in]{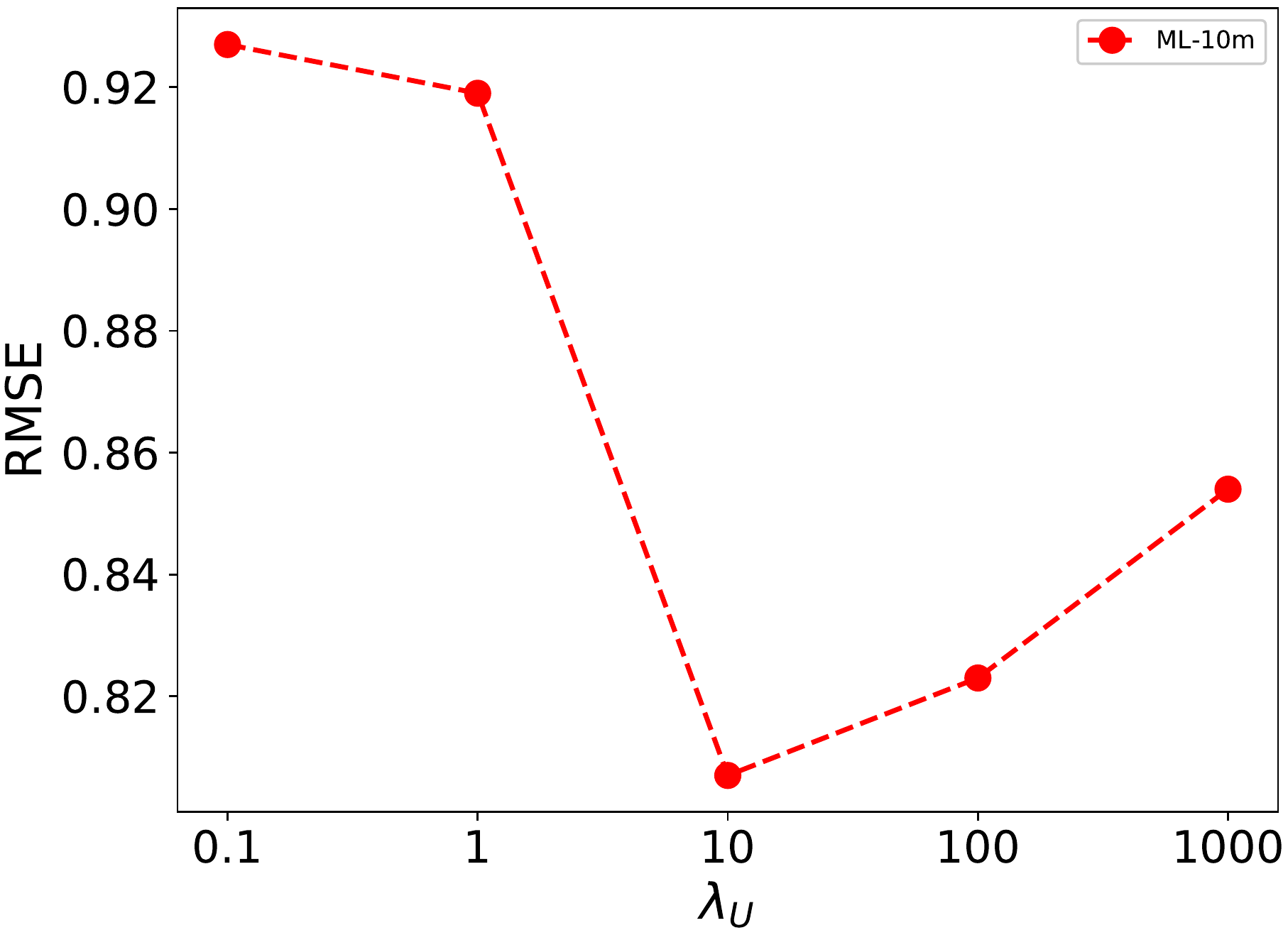}
\end{minipage}
}%
\subfigure{
\begin{minipage}[t]{0.48\linewidth}
\centering
\includegraphics[width=1.7in]{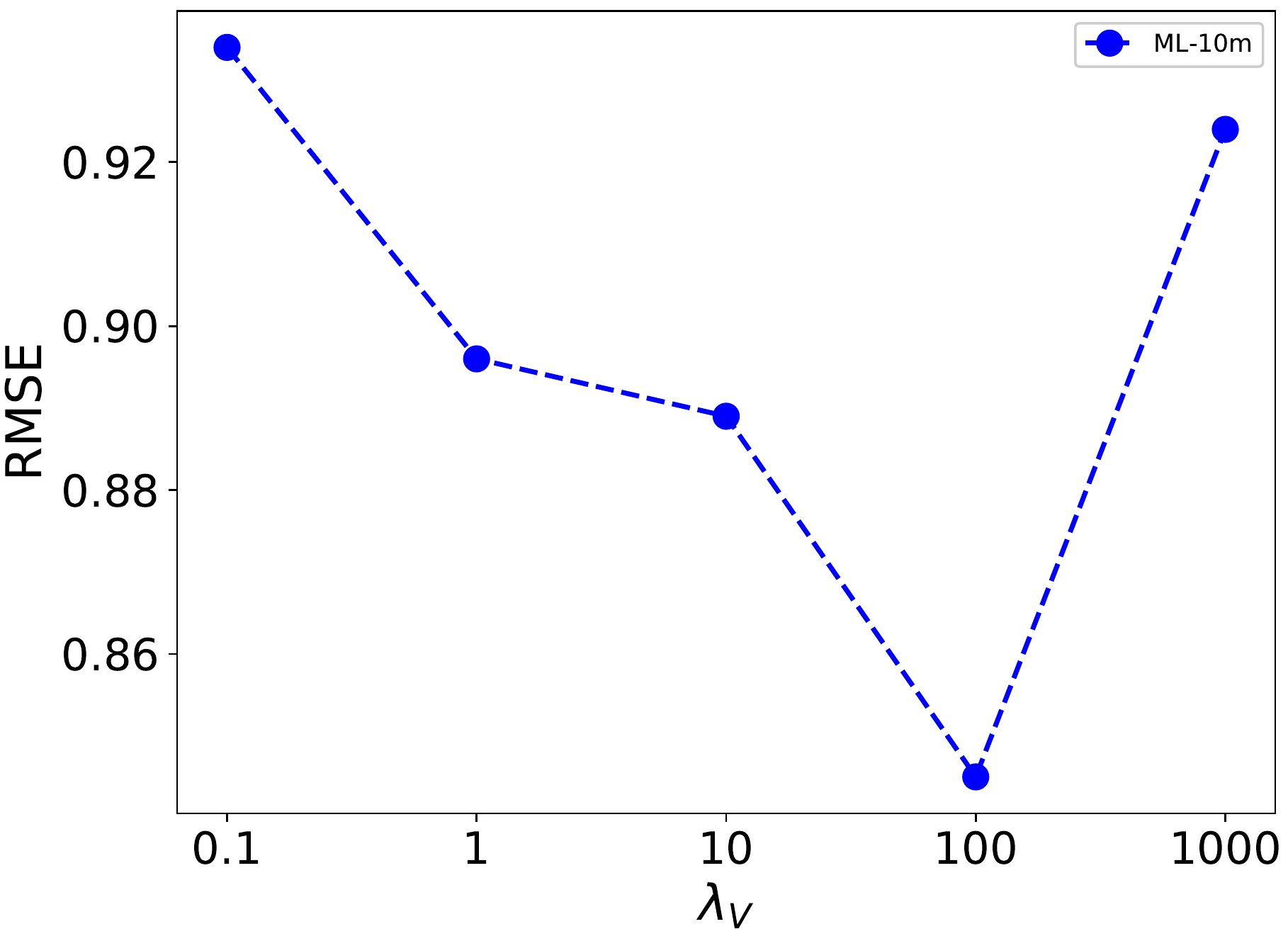}
\end{minipage}
}%

\subfigure{
\begin{minipage}[t]{0.48\linewidth}
\centering
\includegraphics[width=1.7in]{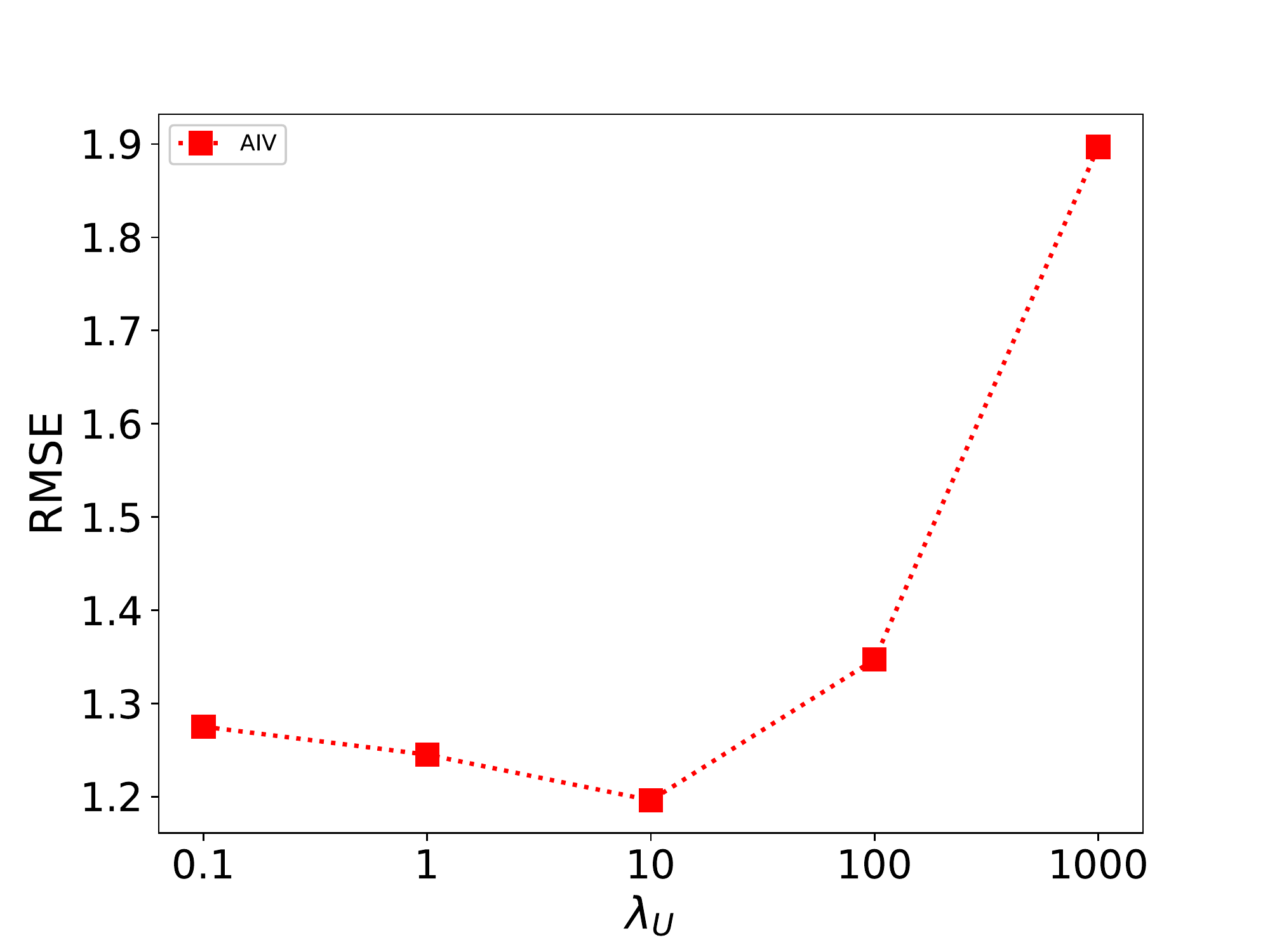}
\end{minipage}
}%
\subfigure{
\begin{minipage}[t]{0.49\linewidth}
\centering
\includegraphics[width=1.7in]{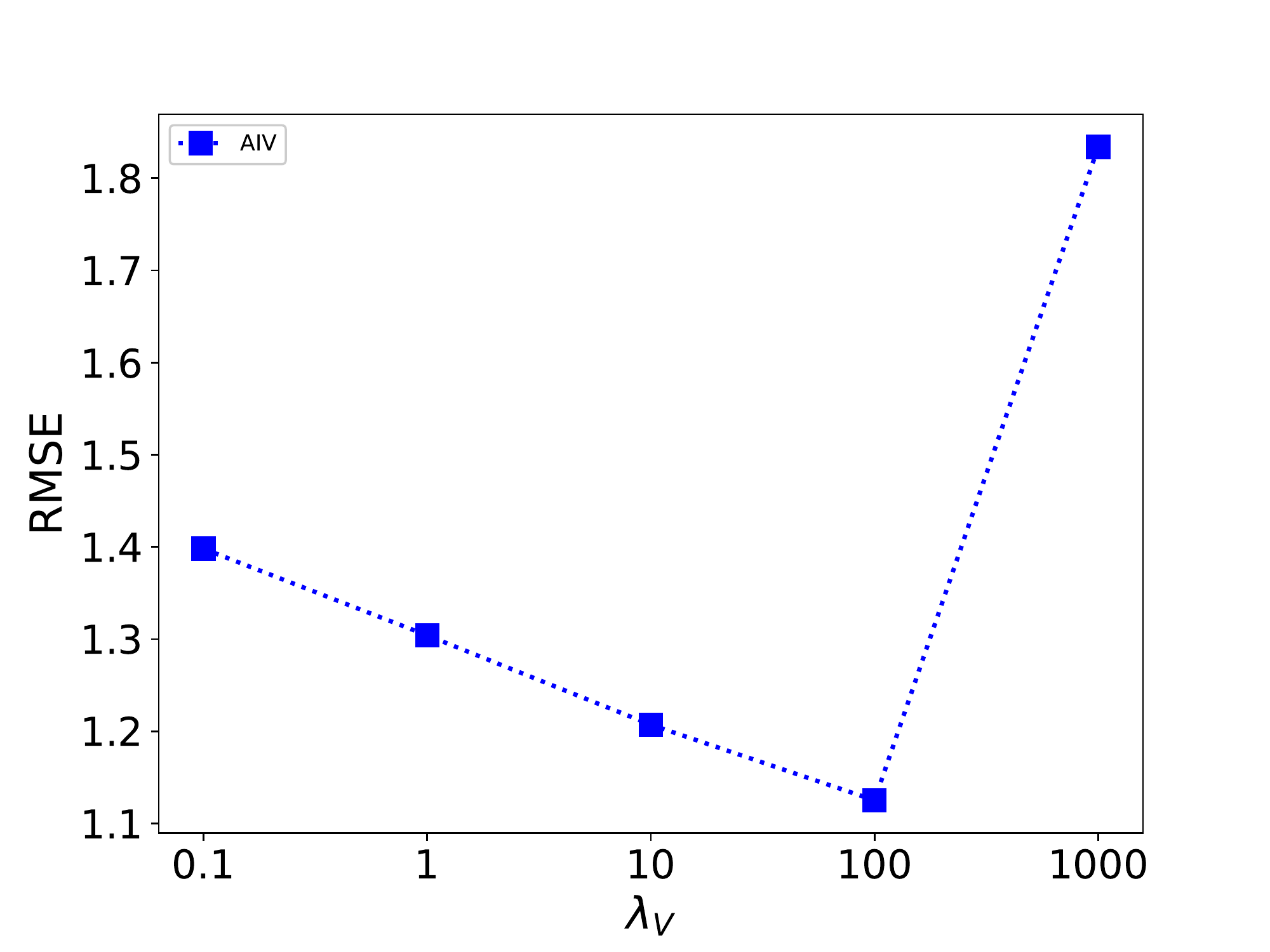}
\end{minipage}
}%
\centering
\caption{Parameter analysis of $\lambda_U$ and $\lambda_V$ on three datasets of the proposed VRConvMF model.}
\label{figure9}
\end{figure}

As for visual feature extraction, firstly, imagenet dataset is applied to pre-training VGG19 network. Then, we utilize the weights of the pre-trained VGG19 network as initial weights so that the network can be converged in a short time and improve the performance of visual feature extraction. Similar to~\cite{K.Simonyan}, batch size, momentum and dropout rate are set to 256, 0.9 and 0.5 respectively. Different from traditional VGG19 network, none of fully connected layers is used when extracting visual features. VGG19 network contains 5 max-pooling layers from shallow to deep. The last 4 max-pooling layer feature maps are taken after pooling operation. Global average pooling method is applied to reduce the multi-dimensional vectors to one-dimensional vectors. The specific dimensions are $\big(1\times1\times128\big)$, $\big(1\times1\times256\big)$, $\big(1\times1\times512\big)$, $\big(1\times1\times512\big)$ for the second, third, forth, and fifth max-pooling layers respectively. We cascade each part of these four layers with the obtained textual feature vector. Obviously, the comprehensive feature for each movie $j$ consists of textual feature and visual feature. Finally, the comprehensive feature vectors are put into the projection layer, fix their dimension to 50, and choose the user latent vectors with the same dimension. For the confidence factor, $\alpha$ is set to 0.3 in Eq.~(\ref{eq12}). The $Grid$ $Search$ method is utilized to find the best performing values of hyper-parameter ($\lambda_U$, $\lambda_V$) of each model and the results are summarized in TABLE~\ref{table9}. Specifically, the detailed parameter optimization experiment processes of the proposed VRConvMF model are shown in Fig.~\ref{figure9}. Since the UserCF is the basic collaborative filtering method without PMF, the parameters are not listed. Besides, the parameters of ConvMF+ are the same with that of ConvMF. Since $\lambda_U$ and $\lambda_V$ are completely independent, the corresponding optimal solutions can be found respectively.

\begin{table}
\caption{Parameter setting of $\lambda_U$ and $\lambda_V$.}\label{table1}
\begin{spacing}{1}
\centering
\label{table9}
\begin{tabular}{c c c c c c c}
\toprule[1.5pt]
\multirow{2}{*}{Model}&\multicolumn{2}{c}{ML-1m}&\multicolumn{2}{c}{ML-10m}&\multicolumn{2}{c}{AIV}\\
\cline{2-7}
&$\lambda_U$&$\lambda_V$&$\lambda_U$&$\lambda_V$&$\lambda_U$&$\lambda_V$\\

\bottomrule[1.5pt]
PMF&0.01&10000&10&100&0.1&0.1\\
\hline
ConvMF&100&10&10&100&1&100\\
\hline
RConvMF&100&10&10&100&1&100\\
\hline
VConvMF&100&10&10&100&10&100\\
\hline
VRConvMF&100&10&10&100&10&100\\
\toprule[1.5pt]
\end{tabular}
\end{spacing}
\end{table}

\subsubsection{\textbf{Evaluation Metric}}
Before training, we randomly split each dataset into a training set \big(80\%\big), a test set \big(10\%\big) and a validation set \big(10\%\big). Then, in order to validate the performance of the proposed VConvMF model, we select the root mean squared error \big(RMSE\big) as the evaluation metric. RMSE can be directly related to an objective function of conventional rating prediction model as follows:
\begin{equation}
RMSE =\sqrt{\frac{\sum\limits_{i = 1}^m\sum\limits_{j = 1}^n\big(r_{ij}-\widehat{r_{ij}}\big)^2}{\big|N\big|}}, \label{eq23}
\end{equation}
where $\big|N\big|$ is the number of test ratings. We set the number of iterations to 30. An average value of 5 repeated experiments are performed as the final result to reduce the random error.

\subsection{Compared Schemes}

\begin{itemize}
\item $\textbf{UserCF}$~\cite{J.L.Herlocker}: UserCF is a classical recommendation algorithm in collaborative filtering, and predicts missing ratings of target users in terms of ratings of similar users.

\item $\textbf{PMF}$~\cite{Salakhutdinov:2008}: Probabilistic matrix factorization is a basic rating prediction that only uses ratings for collaborative filtering.

\item $\textbf{ConvMF}$~\cite{D.Kim}: Convolutional matrix factorization is another context-aware recommendation model, which combines CNN and PMF to enhance the accuracy of rating prediction.
\item $\textbf{ConvMF+}$: The ConvMF+ method considers the confidence mechanism on the basis of ConvMF.
\item $\textbf{RConvMF}$: Recurrent convolutional matrix factorization is a context-aware recommendation model, which combines RCNN and PMF to improve the accuracy of rating prediction.
\item $\textbf{VConvMF}$: Visual convolutional matrix factorization is the preliminary model, which we have proposed above. VConvMF combines both textual features and multi-level visual features by convolutional neural networks.

\end{itemize}

\begin{table*}
\centering
\caption{RMSE of overall test sets.}\label{table2}
\begin{spacing}{1}
\begin{tabular}{p{2.88cm}<{\centering}|p{1.3cm}<{\centering}|p{1.3cm}<{\centering}|p{1.3cm}<{\centering}|p{1.4cm}<{\centering}|p{1.5cm}<{\centering}|p{1.5cm}<{\centering}|p{1.7cm}<{\centering}|p{1.3cm}<{\centering}}
\bottomrule[1.5pt]
\diagbox[dir=NW]{Datasets}{Methods}&UserCF&PMF&ConvMF&ConvMF+
&RConvMF&VConvMF&VRConvMF&Improved\\
\hline
ML-1m&1.5763&1.0129&0.8560&0.8494&0.8453&0.8361&\textbf{0.8201}&4.194\%\\
ML-10m&1.6266&0.9572&0.7966&0.7896&0.7840&0.7756&\textbf{0.7545}&5.285\%\\
AIV&1.3780&1.4322&1.1326&1.1297&1.1125&1.0892&\textbf{1.0373}&8.414\%\\
\toprule[1.5pt]
\end{tabular}
\end{spacing}
\end{table*}

\subsubsection{\textbf{Overall Performance}}
TABLE\ref{table2} shows the overall rating prediction error of the proposed VRConvMF and other six schemes on three real-world datasets. It shows that compared with the traditional schemes (UserCF and PMF), the context-aware recommendation (ConvMF, RConvMF and VConvMF) can achieve a large improvement and the proposed VRConvMF outperforms other schemes. The comparison of the PMF and ConvMF schemes intuitively illustrates the improvement of the accuracy of the recommender systems by deep context understanding. The comparison of the ConvMF and ConvMF+ methods in TABLE~\ref{table2} shows that the confidence mechanism can slightly improve the accuracy of rating prediction. In addition, by comparing the RConvMF model with the ConvMF model, the RCNN model is more effective than the CNN model in textual feature extraction. Finally, as for the proposed VRConvMF model, the visual features are considered to further improve the performance of the RConvMF model. Moreover, from the comparison of the results of the RConvMF model and the VConvMF model, the effect of textual feature is lower than that of visual feature in improving the context-aware recommendation performance. Specifically, compared with the ConvMF, the improvements of VRConvMF are 4.194\%, 5.285\% and 8.414\% for ML-1m, ML-10m and AIV datasets respectively. Moreover, the improvement for the AIV is much more than that for the ML-1m and ML-10m datasets. The results illustrate that the proposed VRConvMF achieves a greater improvement when the dataset has higher sparsity. TABLE~\ref{table2} shows that VRConvMF model has smaller RMSE than VConvMF. Therefore, RCNN is more competent in textual feature extraction than CNN in context-aware recommendations.

\subsubsection{\textbf{Impact of Contextual Information}}
\begin{figure}[htbp]
  \small
  \centering
  \includegraphics[width=3.6in]{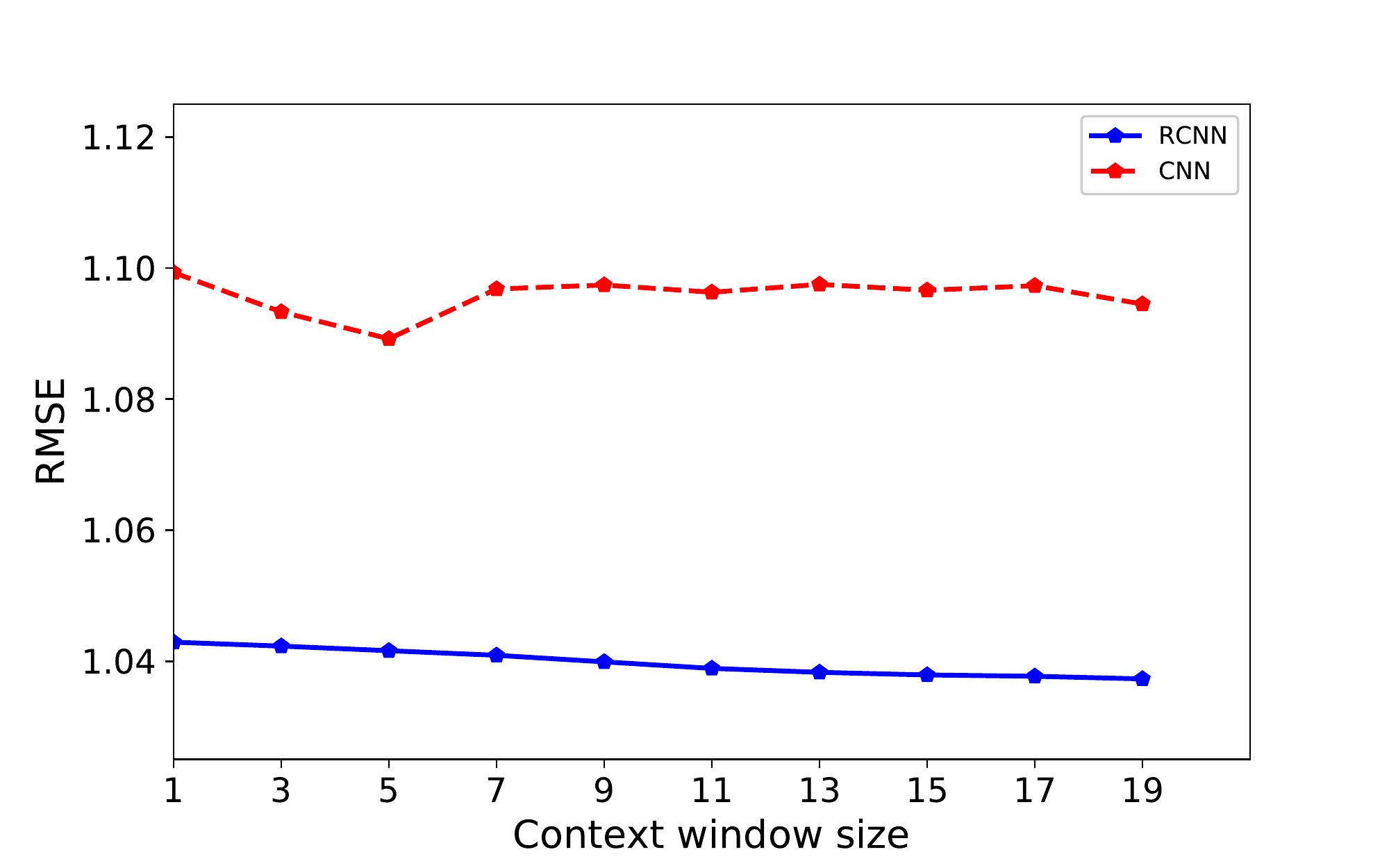}
  \caption{Comparison of different context window sizes of recurrent structure~(AIV).}
  \label{figure4}

\end{figure}

\begin{figure}[htbp]
  \small
  \centering
  \includegraphics[width=3.6in]{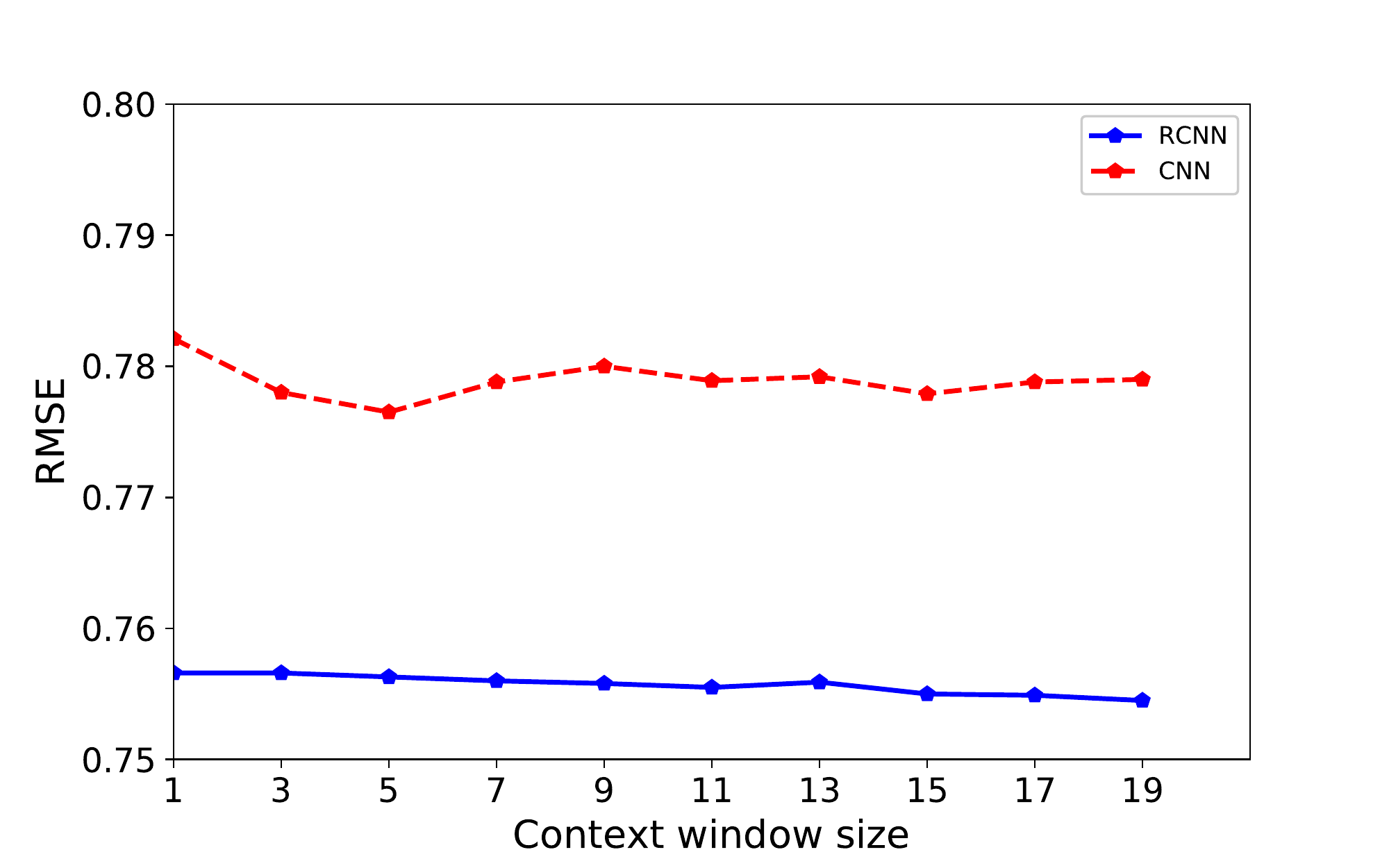}
  \caption{Comparison of different context window sizes of recurrent structure~(ML-10m).}
  \label{figure5}

\end{figure}

We further conduct experiments to analyze the performance of RCNN and CNN with different context window sizes among two of three datasets (ML-10m and AIV) since ML-1m is a subset of ML-10m. Specifically, odd context window sizes are chosen from 1 to 19 to capture the different contextual information. Small context window sizes may lose a lot of textual semantic information at a long distance, while large context window sizes will lead to data sparsity. Fig.~\ref{figure4} and Fig.~\ref{figure5} show the results of the VRConvMF model when considering contextual information, which is captured by RCNN and CNN with different context window sizes respectively. Obviously, RCNN models outperform CNN models under all different context window sizes. Specifically, the CNN models can achieve the best performance in capturing the contextual information when the window size is set to 5. However, the effect of the context window size on the RCNN models is extremely limited, since the recurrent structure in RCNN can already retain sufficient contextual information.


\subsubsection{\textbf{Impact of Different Visual Features}}
\begin{figure}[htbp]
  \centering
  \includegraphics[width=3.8in]{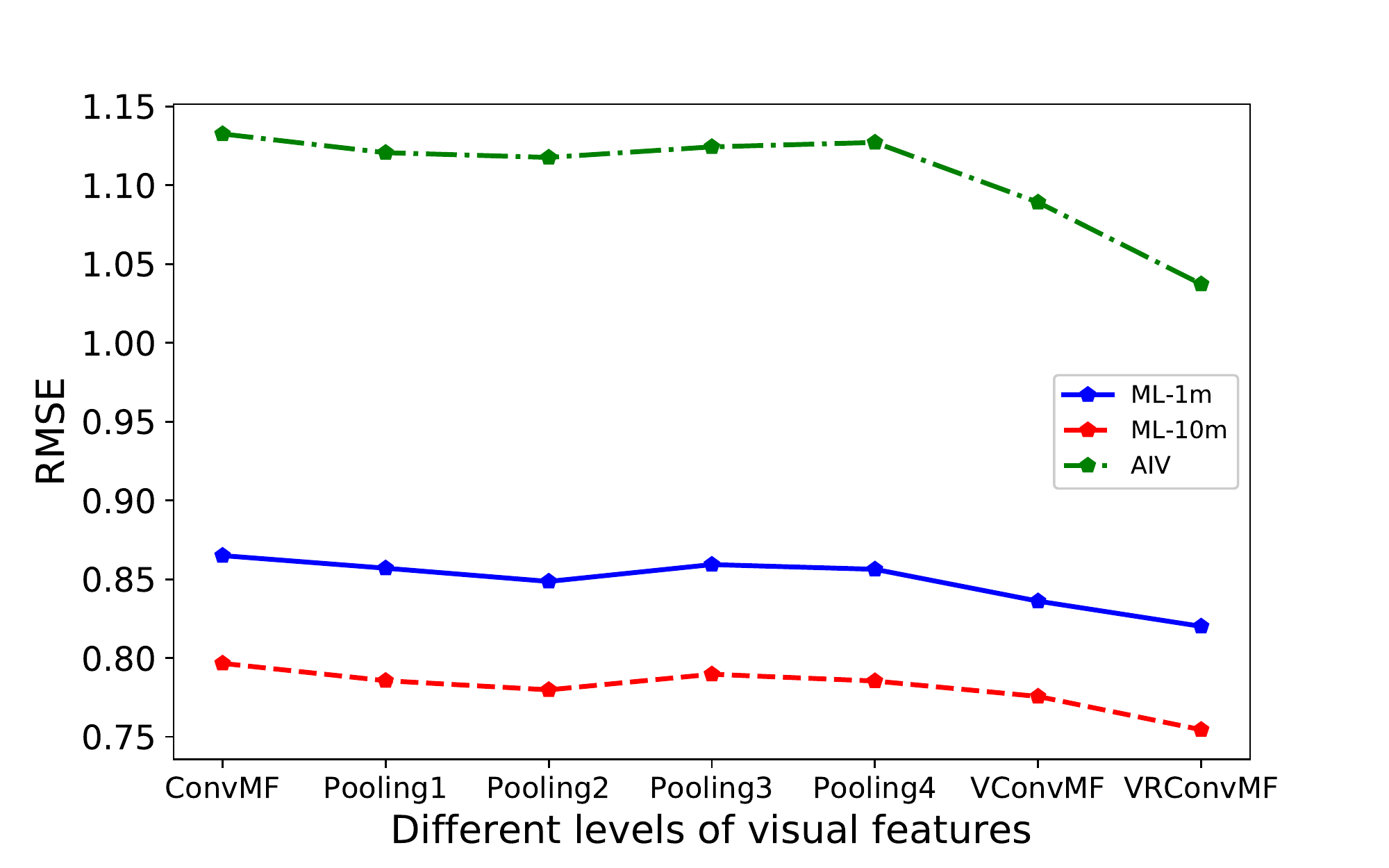}
  \caption{Comparison of the effects of visual features extracted from different convolutional layers on three real-world datasets.}
  \label{figure6}

\end{figure}

We conduct extensive experiments to analyze the impact of single-level visual features at different convolutional layers on the performance of VRConvMF. Fig.~\ref{figure6} shows the improvement of the recommender system when considering the visual features from different levels. Pooling 1 to 4 in Fig.~\ref{figure6} represent the visual features of 4 pooling layers from shallow to deep in the VGG19 network respectively. These single-layers of visual information are combined with RConvMF respectively. It shows that the single-level visual features can reduce the RMSE of the rating prediction to a certain extent. The visual features which are extracted from the mid-level such as pooling 2 tend to outperform those extracted from other pooling layers. Fig.~\ref{figure6} shows that compared with other schemes, the proposed VRConvMF model achieves the best performance as it adds multi-layer visual features extracted from VGG19 on the basis of RCNN as textual feature extraction. Furthermore, the experimental results show that the multi-level visual features can greatly improve the context-aware recommendation performance.
\section{Conclusions and Future Work}\label{sec6}

In this paper, we propose a probabilistic matrix factorization based recommendation scheme called visual recurrent convolutional matrix factorization (VRConvMF). The proposed VRConvMF scheme utilizes the textual and multi-level visual features extracted from the descriptive texts and posters respectively to alleviate the sparsity problem. We implement the proposed VRConvMF scheme and conduct extensive experiments on three commonly used real world datasets. The experimental results illustrate that the proposed VRConvMF scheme outperforms the existing schemes.

In the future work, we will crawl corresponding trailers information of movies directly and transfer them into textual information. Since for visual features, more useful information will appear in the trails. Besides, we also intend to consider the user attributes (such as gender, age and occupations) to improve the rating prediction accuracy.

\bibliography{IEEEtran}

\begin{thebibliography}{10}
\providecommand{\url}[1]{#1}
\csname url@samestyle\endcsname
\providecommand{\newblock}{\relax}
\providecommand{\bibinfo}[2]{#2}
\providecommand{\BIBentrySTDinterwordspacing}{\spaceskip=0pt\relax}
\providecommand{\BIBentryALTinterwordstretchfactor}{4}
\providecommand{\BIBentryALTinterwordspacing}{\spaceskip=\fontdimen2\font plus
\BIBentryALTinterwordstretchfactor\fontdimen3\font minus
  \fontdimen4\font\relax}
\providecommand{\BIBforeignlanguage}[2]{{%
\expandafter\ifx\csname l@#1\endcsname\relax
\typeout{** WARNING: IEEEtran.bst: No hyphenation pattern has been}%
\typeout{** loaded for the language `#1'. Using the pattern for}%
\typeout{** the default language instead.}%
\else
\language=\csname l@#1\endcsname
\fi
#2}}
\providecommand{\BIBdecl}{\relax}
\BIBdecl

\bibitem{Aixin:2021}
X.~Ai, H.~Chen, K.~Lin, Z.~Wang, and J.~Yu, ``{Nowhere to Hide: Efficiently
  Identifying Probabilistic Cloning Attacks in Large-Scale RFID Systems},''
  \emph{IEEE Transactions on Information Forensics and Security}, vol.~16, pp.
  714--727, 2021.

\bibitem{Lin:2021}
K.~Lin, H.~Chen, N.~Yan, Z.~Li, J.~Li, and N.~Jiang, ``{Fast and Reliable
  Missing Tag Detection for Multiple-Group RFID System},'' \emph{IEEE
  Transactions on Industrial Informatics, DOI: 10.1109/TII.2021.305895}, 2021.

\bibitem{G.Adomavicius}
G.~Adomavicius and A.~Tuzhilin, ``Toward the next generation of recommender
  systems: A survey of the state-of-the-art and possible extensions,''
  \emph{IEEE Transactions on Knowledge and Data Engineering}, vol.~17, no.~6,
  pp. 734--749, June 2005.

\bibitem{D.W.Oard}
D.~W. Oard and J.~Kim, ``Implicit feedback for recommender systems,'' in
  \emph{Proc. of AAAI}, 1998, pp. 81--83.

\bibitem{X.Kong}
X.~Kong, F.~Xia, J.~Wang, A.~Rahim, and S.~K.Das, ``Time-location-relationship
  combined service recommendation based on taxi trajectory data,'' \emph{IEEE
  Transactions on Industrial Informatics}, vol.~13, no.~3, pp. 1202--1212,
  2017.

\bibitem{Xiao:2020}
Y.~Xiao, Q.~Pei, L.~Yao, S.~Yu, and X.~Wang, ``An enhanced probabilistic
  fairness-aware group recommendation by incorporating social activeness,''
  \emph{Journal of Network and Computer Applications}, vol. 156, pp. 1--17,
  2020.

\bibitem{Li:2021}
Z.~Li, H.~Chen, K.~Lin, V.~Shakhov, L.~Shi, and J.~Yu, ``{From edge data to
  recommendation: A double attention-based deformable convolutional network},''
  \emph{Peer-to-Peer Networking and Applications, DOI:
  10.1007/s12083-020-01037-7}, 2021.

\bibitem{Chen:2021}
H.~Chen, S.~Wang, N.~Jiang, Z.~Li, N.~Yan, and L.~Shi, ``{Trust-aware
  Generative Adversarial Network with Recurrent Neural Network for
  Recommender},'' \emph{International Journal of Intelligent Systems}, vol.~36,
  pp. 778--795, 2021.

\bibitem{G.Ling}
G.~Ling, M.~R. Lyu, and I.~King, ``Ratings meet review, a combined approach to
  recommend,'' in \emph{Proc. of ACM RecSys}.\hskip 1em plus 0.5em minus
  0.4em\relax Foster City, Silicon Valley, CA, USA: NIPS, October 2014, pp.
  105--112.

\bibitem{Y.Moshfeghi}
Y.~Moshfeghi, B.~Piwowarski, and J.~M.Jose, ``Handling data sparsity in
  collaborative filtering using emotion and semantic based features,'' in
  \emph{Proc. of ACM SIGIR}, New York, NY, USA, 2011, pp. 625--634.

\bibitem{C.Wang}
C.~Wang and D.~M. Blei, ``{Collaborative topic modeling for recommending
  scientific articles},'' in \emph{Proc. of ACM SIGKDD}, 2011, pp. 448--456.

\bibitem{M.Fu}
M.~Fu, H.~Qu, Z.~Yi, L.~Lu, and Y.~Liu, ``A novel deep learning-based
  collaborative filtering model for recommendation system,'' \emph{IEEE
  Transactions on Cybernetics}, vol.~49, no.~3, pp. 1084--1096, March 2019.

\bibitem{F.Xia1}
F.~Xia, J.~Liu, H.~Nie, Y.~Fu, L.~Wan, and X.~Kong, ``Random walks: a review of
  algorithms and applications,'' \emph{IEEE Transactions on Emerging Topics in
  Computational Intelligence}, vol.~4, no.~2, pp. 95--107, 2020.

\bibitem{F.Xia2}
F.~Xia, H.~Liu, I.~Lee, and L.~Cao, ``Scientific article recommendation:
  exploiting common author relations and historical preferences,'' \emph{IEEE
  Transactions on Big Data}, vol.~2, no.~2, pp. 101--112, 2016.

\bibitem{H.Wang}
H.~Wang, N.~Wang, and D.-Y. Yeung, ``Collaborative deep learning for
  recommender systems,'' in \emph{Proc. of ACM SIGKDD}, Sydney, NSW, Australia,
  August 2015, pp. 1235--1244.

\bibitem{Salakhutdinov:2008}
R.~Salakhutdinov and A.~Mnih, ``Probabilistic matrix factorization,'' in
  \emph{Proc. of NIPS}, 2008, pp. 1257--1264.

\bibitem{Y.Kim}
Y.~Kim, ``Convolutional neural networks for sentence classification,'' in
  \emph{Proc. of EMNLP}, Doha, Qatar, October 2014, pp. 1746--1751.

\bibitem{J.Pennington}
J.~Pennington, R.~Socher, and C.~D. Manning, ``Glove: Global vector for word
  representation,'' in \emph{Proc. of EMNLP}, 2014, pp. 1532--1543.

\bibitem{L.Zhao}
L.~Zhao, Z.~Lu, S.~J. Pan, and Q.~Yang, ``Matrix factorization+ for movie
  recommendation,'' in \emph{Proc. of IJCAI}, 2016, pp. 3945--3951.

\bibitem{S.Lai}
S.~Lai, L.~Xu, K.~Liu, and J.~Zhao, ``Recurrent convolutional neural networks
  for text classification,'' in \emph{Proc. of AAAI}, Sydney, NSW, Australia,
  August 2015, pp. 2267--2273.

\bibitem{D.Kim}
D.~Kim, C.~Park, J.~Oh, S.~Lee, and H.~Yu, ``Convolutional matrix factorization
  for document context-aware recommendation,'' in \emph{Proc. of ACM RecSys},
  Boston, MA, USA, September 2016, pp. 233--240.

\bibitem{H.Chen}
H.~Chen, J.~Fu, L.~Zhang, S.~Wang, K.~Lin, and L.~Shi, ``Deformable
  convolutional matrix factorization for document context-aware recommendation
  in social networks,'' \emph{IEEE Access}, vol.~7, pp. 66\,347--66\,357, May
  2019.

\bibitem{P.Covington}
P.~Covington, J.~Adams, and E.~Sargin, ``Deep neural networks for youtube
  recommendations,'' in \emph{Proc. of ACM RecSys}, Boston, MA, USA, 2016, pp.
  191--198.

\bibitem{S.Duan}
S.~Duan, D.~Zhang, Y.~Wang, L.~Li, and Y.~Zhang, ``Jointrec: A
  deep-learning-based joint cloud video recommendation framework for mobile
  iot,'' \emph{IEEE Internet of Things Journal}, vol.~7, no.~3, pp. 1655--1666,
  March 2020.

\bibitem{Z.Huang}
Z.~Huang, J.~Tang, G.~Shan, J.~Ni, Y.~Chen, and C.~Wang, ``An efficient
  passenger-hunting recommendation framework with multitask deep learning,''
  \emph{IEEE Internet of Things Journal}, vol.~6, no.~5, pp. 7713--7731,
  October 2019.

\bibitem{Y.Koren1}
Y.~Koren, ``Collabortive filtering with temporal dynamics,''
  \emph{Communications of the ACM}, vol.~53, no.~4, pp. 89--97, 2010.

\bibitem{H.Stormer}
H.~Stormer, ``Improving e-commerce recommender systems by the identification of
  seasonal products,'' in \emph{Proc. of AAAI}, 2007, pp. 92--99.

\bibitem{A.Q.Macedo}
A.~Q. Macedo, L.~B. Marinho, and R.~L.~T. Santos, ``Context-aware event
  recommendation in event-based social networks,'' in \emph{Proc. of ACM
  Recsys}, Vienna, Austria, September 2015, pp. 123--130.

\bibitem{S.Purushotham}
S.~Purushotham, Y.~Liu, and C.-C.~J. Kuo, ``Collaborative topic regression with
  social matrix factorization for recommendation systems,'' in \emph{Proc. of
  ICML}, Edinburgh, Scotland, UK, 2012, pp. 759--766.

\bibitem{J.McAuley}
J.~McAuley and J.~Leskovec, ``Hidden factors and hidden topics: understanding
  rating dimensions with review text,'' in \emph{Proc. of ACM RecSys}, New
  York, NY, USA, 2013, pp. 165--172.

\bibitem{H.Negar}
N.~Hariri, B.~Mobasher, and R.~Burke, ``Context-aware music recommendation
  based on latent topic sequential patterns,'' in \emph{Proc. of ACM RecSys},
  Dublin, Ireland, September 2012, pp. 131--138.

\bibitem{R.Cheng}
R.~Cheng and B.~Tang, ``A music recommendation system based on acoustic
  features and user personalities,'' in \emph{Proc. of Pacific-Asia Conference
  on Knowledge Discovery and Data Mining}, 2016, pp. 203--213.

\bibitem{S.Oramas}
S.~Oramas, O.~Nieto, M.~Sordo, and X.~Serra, ``A deep multimodal approach for
  cold-start music recommendation,'' in \emph{Proc. of DLRS}, Como, Italy, June
  2017, pp. 32--37.

\bibitem{Y.Fan}
Y.~Fan, Y.~Wang, H.~Yi, and B.~Liu, ``Movie recommendation based on visual
  features of trailers,'' in \emph{Proc. of International Conference on
  Innovative Mobile and Internet Services in Ubiquitous Computing}, Torino,
  Italy, July 2017, pp. 242--252.

\bibitem{Vinyals:2015}
O.~Vinyals, A.~Toshev, S.~Bengio, and D.~Erhan, ``Show and tell: a neural image
  caption generator,'' in \emph{Proc. of IEEE CVPR}, 2015, pp. 3156--3164.

\bibitem{S.A.-E.-H.}
S.~Abu-El-Haija, N.~Kothari, J.~Lee, P.~Natsev, G.~Toderici, B.~Varadarajan,
  and S.~Vijayanarasimhan, ``Youtube-8m: A large-scale video classification
  benchmark,'' \emph{arXiv preprint arXiv:1609.08675}, September 2016.

\bibitem{Y.Koren}
Y.~Koren, R.~Bell, and C.~Volinsky, ``Matrix factorization techniques for
  recommender systems,'' \emph{Computer}, vol.~42, no.~8, pp. 42--49, August
  2009.

\bibitem{T.Mikolov}
T.~Mikolov, K.~Chen, G.~Corrado, and J.~Dean, ``Efficient estimation of word
  representations in vector space,'' in \emph{Proc. of IJCAI}, 2013.

\bibitem{K.Simonyan}
K.~Simonyan and A.~Zisserman, ``Very deep convolutional networks for
  large-scale image recognition,'' \emph{Computer Science}, pp. 770--778, 2014.

\bibitem{W.Yu}
W.~Yu, K.~Yang, H.~Yao, X.~Sun, and P.~Xu, ``Exploiting the complementary
  strengths of multi-layer cnn features for image retrieval,''
  \emph{Neurocomputing}, vol. 237, pp. 235--241, July 2017.

\bibitem{J.L.Herlocker}
J.~L. Herlocker, J.~A. Konstan, A.~Borchers, and J.~Riedl, ``An algorithmic
  framework for performing collaborative filtering,'' in \emph{Proc. of ACM
  SIGIR}, Berkley, CA, USA, October 1999, pp. 230--237.

\end{thebibliography}

\bibliographystyle{IEEEtran}
\begin{IEEEbiography}[{\includegraphics[width=1in,height=1.25in,clip,keepaspectratio]{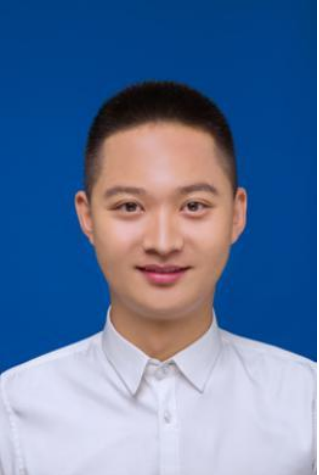}}]{Zhu Wang}
received the B.E. degree in Electrical Engineering and Automation from University of Jinan, China, in 2017. He is currently a graduate student in the College of Control Science and Engineering, China University of Petroleum, China. His research interest
is in the field of recommender system.
\end{IEEEbiography}

\begin{IEEEbiography}[{\includegraphics[width=1in,height=1.25in,clip,keepaspectratio]{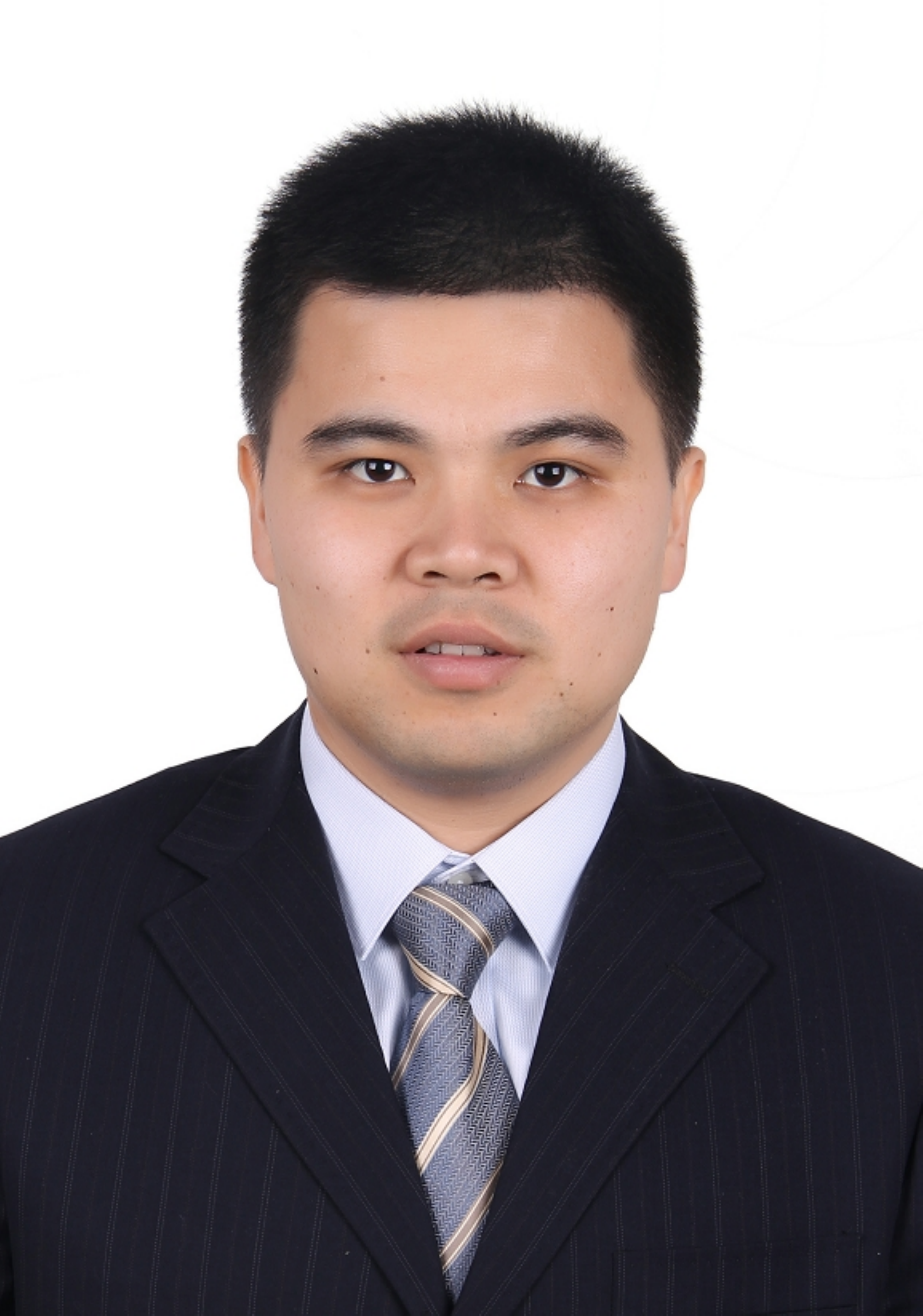}}]{Honglong Chen}
received the M.E. degree in Control Science and Engineering from Zhejiang University, China, in 2008, and
the Ph.D degree in computer science from The Hong Kong Polytechnic
University, Hong Kong, in 2012. He was a Postdoctoral Researcher in
the School of CIDSE at Arizona State University from 2015
to 2016. He is currently an Associate Professor and Ph.D supervisor with the
College of Control Science and Engineering, China University of
Petroleum, China. He served as a Guest Editor of IEEE Transactions on Industrial Informatics. His current research interests are in the areas of Internet of Things and recommender systems. He has published more than 70 research papers in prestigious journals and conferences including IEEE TIFS, IEEE TMC, IEEE IoT-J, IEEE TII, IEEE TWC, IEEE INFOCOM, IEEE ICPP, IEEE ICDCS, etc. He is a senior member of IEEE and CCF (China Computer Federation).
\end{IEEEbiography}

\begin{IEEEbiography}[{\includegraphics[width=1in,height=1.25in,clip,keepaspectratio]{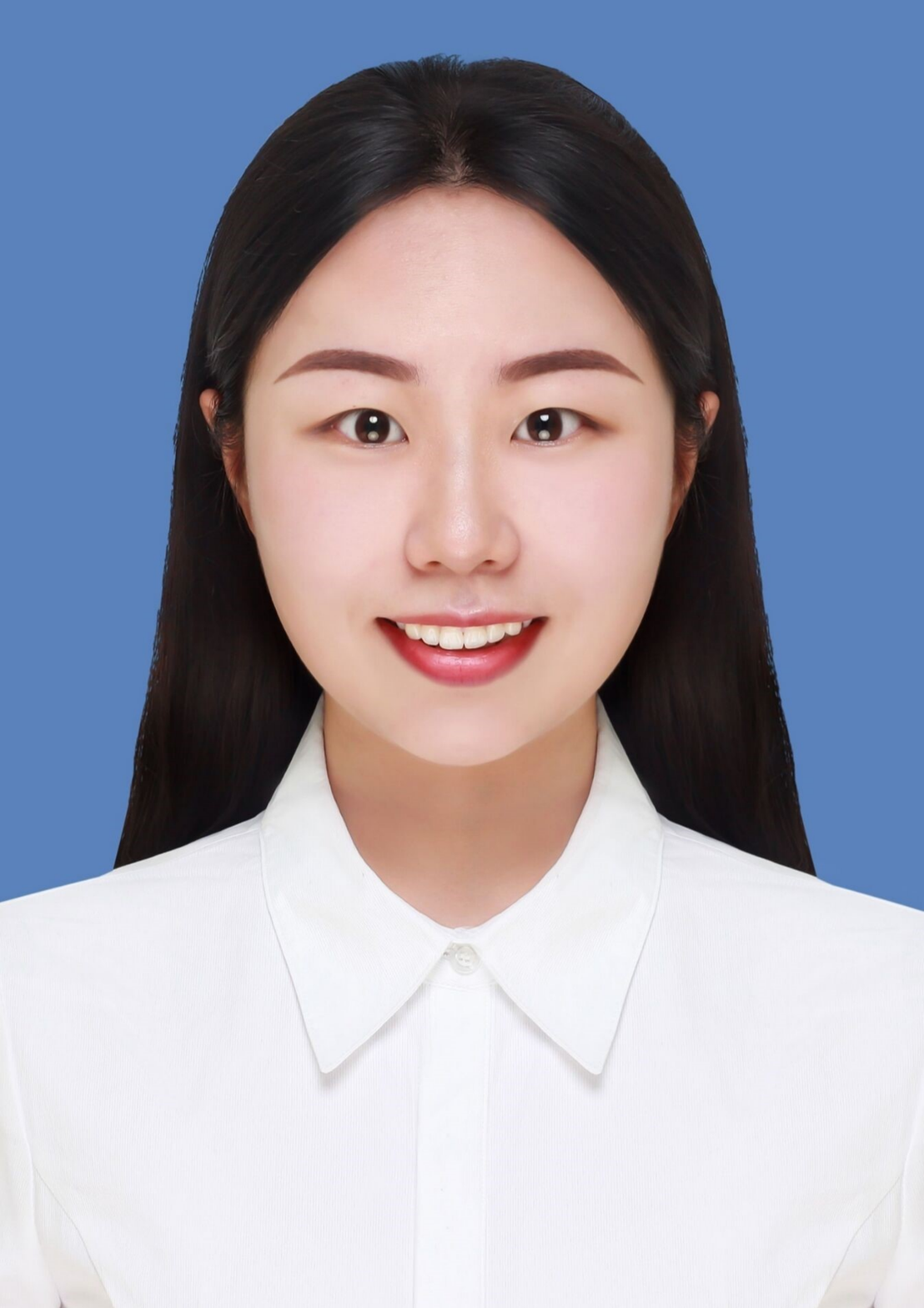}}]{Zhe Li}
received the B.E. degree in Automation from Shandong University of Science and Technology, China, in 2018. She is currently a graduate student in the College of Control Science and Engineering, China University of Petroleum, China. Her research interest is in the field of recommender system.
\end{IEEEbiography}

\begin{IEEEbiography}[{\includegraphics[width=1in,height=1.25in,clip,keepaspectratio]{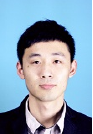}}]{Kai Lin}
received the B.E. degree in Automation from Tianjin Polytechnic University, China, in 2015 and the M.E. degree in Control Engineering from China University of Petroleum, China, in 2019. He is currently pursuing his Ph.D degree in control science and engineering in the College of Control Science and Engineering, China University of Petroleum, China. His current research interests include RFID system and Internet of things.
\end{IEEEbiography}

\begin{IEEEbiography}[{\includegraphics[width=1in,height=1.25in,clip,keepaspectratio]{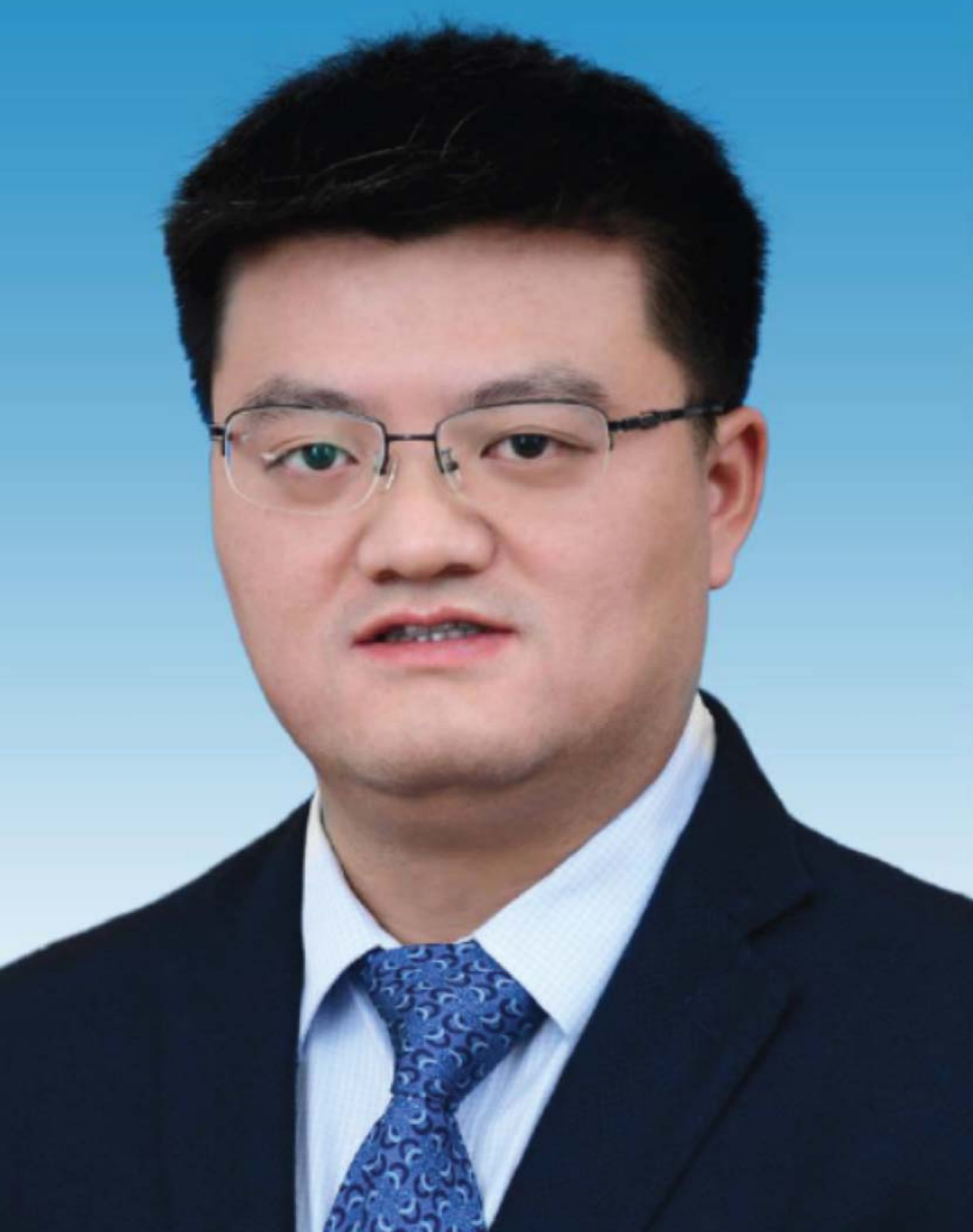}}]{Nan Jiang} received the
Ph.D degree from the Nanjing University
of Aeronautics and Astronautics in 2008. From 2013 to
2014, he was a research scholar in the Complex
Networks and Security Research Lab at Virginia
Tech, Blacksburg, VA, USA. He is currently a
Professor with Department of Internet of Things and
the director of the Intelligent Sensor Networks Lab
at East China Jiaotong University. His current research interests lie in the Internet of Things, Cyber Physical
Systems, and Social Networks. He has published more than 60 papers in international
journals and conferences and served as program chairs and committee for
many international conferences such as CSS 2019, IEEE EUC 2017, ISICA
2015 etc. He is a member of IEEE, a member of ACM, and a senior member of the China Computer Federation (CCF).
\end{IEEEbiography}

\begin{IEEEbiography}[{\includegraphics[width=1in,clip,keepaspectratio]{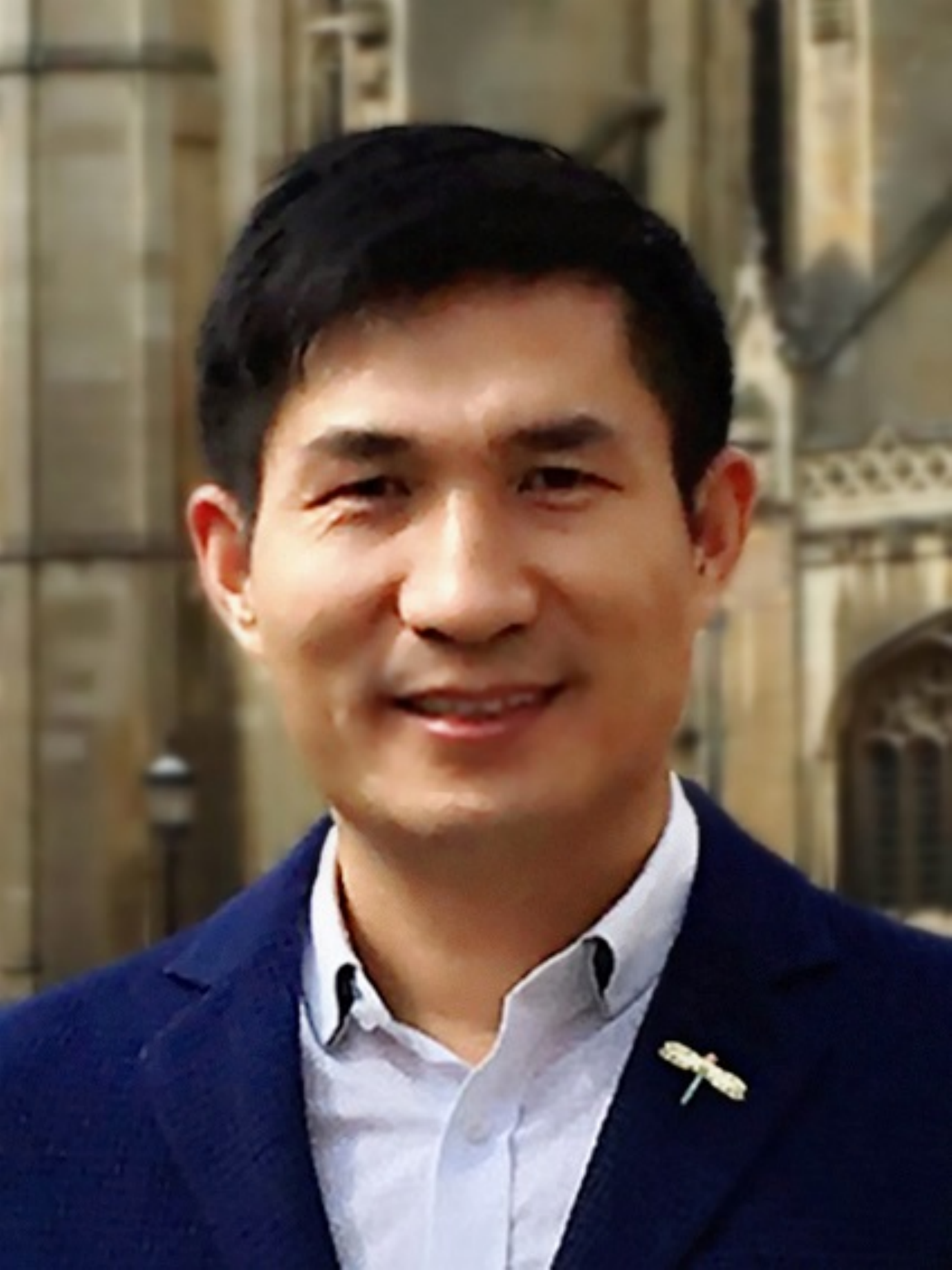}}]{Feng Xia}
 received the BSc and PhD degrees from Zhejiang University, Hangzhou, China. He was a Full Professor and Associate Dean (Research) in School of Software, Dalian University of Technology, China. He is currently an Associate Professor and Discipline Leader in School of Engineering, IT and Physical Sciences, Federation University Australia. Dr. Xia has published 2 books and over 300 scientific papers in international journals and conferences (such as IEEE TAI, TKDE, TBD, TCSS, TNSE, TETCI, TC, TMC, TPDS, TETC, THMS, TVT, TITS, TASE, ACM TKDD, TIST, TWEB, TOMM, WWW, AAAI, SIGIR, CIKM, JCDL, EMNLP, and INFOCOM). His research interests include data science, artificial intelligence, and social computing. He is a Senior Member of IEEE and ACM.
\end{IEEEbiography}
\end{document}